\documentclass[a4paper,fleqn,usenatbib]{mnras}

\usepackage{newtxtext,newtxmath}
\usepackage[T1]{fontenc}
\usepackage{ae,aecompl}

\usepackage{graphicx}	% Including figure files
\usepackage{amsmath}	% Advanced maths commands
\usepackage{amssymb}	% Extra maths symbols
\usepackage{color}

\newcommand{\kms}{km\,s$^{-1}$}
\newcommand{\hbeta}{H$\beta$}
\newcommand{\halpha}{H$\alpha$}
\newcommand{\warmH}{H$_{2}$ (1--0) S(1)}
\newcommand{\bry}{Br$\gamma$}

\title[Multiphase Gas in NGC 5728]{The multiphase gas structure and kinematics in the circumnuclear region of NGC 5728}%\thanks{Based on observations made with ESO Telescopes at the La Silla Paranal Observatory under programme IDs 093.B-0057 and 097.B-0640}}

\author[T. T. Shimizu et al.]{T.~Taro Shimizu$^{1}$\thanks{E-mail: shimizu@mpe.mpg.de},
R.I.~Davies$^{1}$,
D.~Lutz$^{1}$,
L.~Burtscher,$^{2}$,
M.~Lin$^{3}$,
\newauthor
D.~Baron$^{4}$,
R.L.~Davies$^{1}$,
R.~Genzel$^{1}$,
E.K.S.~Hicks$^{5}$,
M.~Koss$^{6}$,
W.~Maciejewski$^{7}$,
\newauthor
F.~M\"uller-S\'anchez$^{8}$,
G.~Orban~de~Xivry$^{9}$,
S.H.~Price$^{1}$,
C.~Ricci$^{10}$,
\newauthor
R.~Riffel$^{11}$,
R.A.~Riffel$^{12}$,
D. Rosario$^{13}$,
M.~Schartmann$^{1}$,
A.~Schnorr-M\"uller$^{11}$,
\newauthor
A.~Sternberg$^{4,15}$,
E.~Sturm$^{1}$,
T.~Storchi-Bergmann$^{11}$,
L.~Tacconi$^{1}$, and
S.~Veilleux$^{14}$
\\
$^{1}$Max-Planck-Institut f\"{u}r extraterrestrische Physik, Postfach 1312, D-85741, Garching, Germany\\
$^{2}$Sterrewacht Leiden, Universiteit Leiden, Niels-Bohr-Weg 2, 2300 CA Leiden, The Netherlands\\
$^{3}$Institute of Astronomy and Astrophysics, Academia Sinica, No.1, Sec. 4, Roosevelt Rd, Taipei 10617, Taiwan\\
$^{4}$Raymond and Beverly Sackler School of Physics \& Astronomy, Tel Aviv University, Ramat Aviv 69978, Israel\\
$^{5}$Department of Physics \& Astronomy, University of Alaska Anchorage, AK 99508-4664, USA\\
$^{6}$Eureka Scientific Inc., 2452 Delmer St. Suite 100, Oakland, CA 94602, USA\\
$^{7}$Astrophysics Research Institute, Liverpool John Moores University, IC2 Liverpool Science Park, 146 Brownlow Hill, L3 5RF, UK\\
$^{8}$Department of Physics and Material Science, University of Memphis, Memphis, TN 38152, USA\\
$^{9}$Space Sciences, Technologies, and Astrophysics Research Institute, Universit\'e de Li\`ege, 4000 Sart Tilman, Belgium\\
$^{10}$N\'ucleo de Astronom\'a de la Facultad de Ingenier\'a, Universidad Diego Portales, Av. Ej\'ercito Libertador 441, Santiago, Chile\\
$^{11}$Departamento de Astronomia, Universidade Federal do Rio Grande do Sul, IF, CP 15051, 91501-970 Porto Alegre, RS, Brazil\\
$^{12}$Departamento de F\'isica, Centro de Ci\^encias Naturais e Exatas, Universidade Federal de Santa Maria, 97105-900 Santa Maria, RS, Brazil\\
$^{13}$Department of Physics, Durham University, South Road, Durham, DH1 3LE, UK\\
$^{14}$Department of Astronomy and Joint Space-Science Institute, University of Maryland, College Park, MD 20742-2421 USA\\
$^{15}$Center for Computational Astrophysics, Flatiron Institute, 162 5th Ave, New York, NY, 10010, USA
}

\date{Accepted XXX. Received YYY; in original form ZZZ}

\pubyear{2019}

\begin{document}
\label{firstpage}
\pagerange{\pageref{firstpage}--\pageref{lastpage}}
\maketitle

\begin{abstract}
We report on our combined analysis of HST, VLT/MUSE, VLT/SINFONI, and ALMA observations of the local Seyfert~2 galaxy, NGC~5728 to investigate in detail the feeding and feedback of the AGN. The datasets simultaneously probe the morphology, excitation, and kinematics of the stars, ionized gas, and molecular gas over a large range of spatial scales (10~pc--10~kpc). NGC~5728 contains a large stellar bar which is driving gas along prominent dust lanes to the inner 1~kpc where the gas settles into a circumnuclear ring. The ring is strongly star forming and contains a substantial population of young stars as indicated by the lowered stellar velocity dispersion and gas excitation consistent with HII regions. We model the kinematics of the ring using the velocity field of the CO~(2--1) emission and stars and find it is consistent with a rotating disk. The outer regions of the disk, where the dust lanes meet the ring, show signatures of inflow at a rate of 1~M$_{\sun}$~yr$^{-1}$. Inside the ring, we observe three molecular gas components corresponding to the circular rotation of the outer ring, a warped disk, and the nuclear stellar bar. The AGN is driving an ionized gas outflow that reaches a radius of 250~pc with a mass outflow rate of 0.08~M$_{\sun}$~yr$^{-1}$ consistent with its luminosity and scaling relations from previous studies. While we observe distinct holes in CO emission which could be signs of molecular gas removal, we find that largely the AGN is not disrupting the structure of the circumnuclear region.
\end{abstract}


%%%%%%%%%%%%%%%%%%%%%%%%%%%%%%%%%%%%%%%%%%%%%%%%%%

%%%%%%%%%%%%%%%%% BODY OF PAPER %%%%%%%%%%%%%%%%%%

\section{Introduction}
Beyond being the some of the most energetic objects in the Universe, active galactic nuclei (AGN) are thought to play an important role in the evolution of their host galaxies. In particular, large scale cosmological simulations require feedback from AGN to reproduce the galaxy population we observe today \citep[e.g.][]{Springel:2005ve,Bower:2006gf,Croton:2006kx,Nelson:2019aa}. More complete knowledge of the processes that fuel AGN and the mechanisms through which they provide feedback is then necessary to understand the path galaxies take from being gas rich and star-forming to gas poor and quiescent. 

To obtain this understanding however requires dedicated observations that cover a large range of wavelengths and spatial scales to reveal the dominant processes. AGN only occupy the very centres of galaxies but both streaming gas fuelling the AGN as well as outflowing gas from AGN driven winds or jets can reach at least several kpc away \citep[e.g.][]{Rupke:2011aa,Cicone:2014ty,Fischer:2013aa,Bae:2017aa,Baron:2018aa,Fischer:2018aa, Herrera-Camus:2018aa,Kang:2018aa,Mingozzi:2018aa,Bischetti:2019em}. Furthermore, this gas does not exist in a single phase but rather consists of a mixture of different phases from cold molecular gas to warm ionized gas to hot X-ray emitting gas. Thus to obtain a full accounting of the direct fuelling and feedback of the AGN will need multiple continuum and emission line tracers to probe each phase of the gas. 

With the rise of integral field unit (IFU) instruments, large radio interferometers and adaptive optics (AO), it is now possible to study simultaneously the distribution and kinematics of both ionized and molecular gas at the same spatial scales in an efficient manner. For nearby galaxies, this translates to a wealth of information that can span tens of pc to hundreds of kpc, and the kinematics of the gas can further be compared to the stellar kinematics through the analysis of stellar absorption lines to reveal non-circular motions such as radial inflow or outflow. Indeed there has been an increased effort in combining mutliwavelength data to obtain a full view of the conditions and dynamics of gas around AGN. Recent examples include NGC 5643 \citep{Alonso-Herrero:2018aa}, ESO 428-G14 \citep{Feruglio:2019aa}, ESO 578-G009 \citep{Husemann:2019aa}, NGC 3393 \citep{Finlez:2018aa}, NGC 1566 \citep{Slater:2019aa}, NGC 2110 \citep{Rosario:2019aa}, and zC400528 \citep{Herrera-Camus:2018aa}. These works primarily paired ALMA observations of CO emission with an optical or NIR IFU observation to measure inflow and outflow velocities and gas masses, gas excitation, and bar driven instabilities. 

The importance of multiphase observations, in particular for studying AGN feedback, was outlined in \citet{Fiore:2017aa} who compiled and reanalysed data probing AGN driven outflows in the molecular, ionized and X-ray emitting phase. Their findings revealed strong correlations between the AGN luminosity and outflow velocity, mass outflow rate, and wind momentum for all phases of the gas. However the scaling relations differ depending on the phase resulting in a changing fraction of outflowing gas in one phase that is dependent on the strength of the AGN. The \citet{Fiore:2017aa} scaling relations also do not extend to lower luminosities that are more representative of the larger population of AGN. It remains to be seen whether they are valid for Seyfert-like AGN with recent work using dust as a tracer of AGN outflows suggests the relations break down \citep{Baron:2019aa}.

Assessing an AGN's impact on its host galaxy requires measurements of important physical properties of the outflowing gas including the radial extent and the electron density. Without spatially resolved data and lines that accurately trace the electron density, estimates of the mass outflow rate and energetics can be off by orders of magnitude. \citet{Revalski:2018aa}, using high spatial resolution \textit{HST} imaging and long slit spectra, showed that while global estimates of outflow properties overall agree with spatially resolved estimates, they come with large uncertainties and highly depend on the assumed geometry and density of the system. Only with precise knowledge of the radial extent and gas density will globally averaged mass outflow rates, kinetic luminosities, and momentum rates come close to those obtained from spatially resolved data.

A parallel and equally important focus of studies of nearby AGN is not how gas is being removed but rather how gas is fuelling the AGN. Many mechanisms have been proposed to extract angular momentum and drive gas to the central supermassive black hole (SMBH) but so far no unique one has been found to explain all observations of AGN. Instead a range of mechanisms can likely fuel AGN that depends on the luminosity and morphology of the host galaxy. For low to moderate luminosity AGN, recent work suggests that gravitational torques produced by non-axisymmetric potentials are efficiently bringing gas down to the inner kiloparsec \citep[e.g.][; however also see \citet{Sormani:2018dj}]{Combes:2003aa, Garcia-Burillo:2005aa, Garcia-Burillo:2012aa}. Hydrodynamic simulations show that gas flowing along a bar settles into a nuclear ring, spatially located in or near the Inner Lindblad Resonance (ILR), but further inflow inner to the ring is small without subsequent dynamical instabilities such as another nested nuclear bar \citep{Combes:1985aa, Piner:1995aa,Regan:2003aa,Maciejewski:2002aa,Maciejewski:2004aa}. Instead of nuclear rings, nuclear spirals can also form and propagate down to the SMBH given a higher gas sound speed \citep{Maciejewski:2002aa, Maciejewski:2004aa} and indeed nuclear spirals have been found in many galaxies hosting an AGN \citep{2002ApJ...569..624P,Martini:2003aa,Prieto:2005aa,Davies:2009aa,Davies:2014aa}. Since cold molecular gas is the fuel for AGN, it makes sense to turn to observations of CO line emission, a primary tracer for molecular gas, to detect and measure the structure and instabilities bringing gas to the SMBH.

\subsection{NGC 5728}\label{sec:ngc5728}
NGC 5728 is a nearby SAB(r)a galaxy ($D = 39$ Mpc, 1\arcsec = 190 pc), with a Compton thick, Seyfert 2 nucleus \citep{Veron:1981aa,Phillips:1983aa}. The bolometric luminosity of the AGN is $1.40\times10^{44}$ ergs s$^{-1}$ where we have converted the absorption corrected 14--195 keV X-ray luminosity from the \textit{Swift} Burst Alert Telescope, $L_{X} = 2.15\times10^{43}$ ergs s$^{-1}$ \citep{Ricci:2017aa}, into $L_{\rm Bol}$ using the relation from \citet{Winter:2012yq}.
\newpage Its primary characteristics include a large stellar bar \citep[$R\approx11$ kpc, P.A. = 33\degr;][]{Schommer:1988aa,Prada:1999aa} which is surrounded by a ring of young stars, a circumnuclear star forming ring \citep[$R\approx800$ pc;][]{Schommer:1988aa, Wilson:1993aa}, and extended ionization cones \citep[$R\approx1.5$ kpc, P.A. =118\degr;][]{Schommer:1988aa, Arribas:1993aa, Wilson:1993aa, Mediavilla:1995aa}. \citet{Rubin:1980aa} was the first to study the ionized gas kinematics finding that while at large radii the gas displayed normal rotation, the central regions showed strong non-circular motion as well as double peaked emission lines. Over the years much discussion ensued over the physical explanation of the non-circular motion and double peaked emission lines with radial outflow \citep{Rubin:1980aa, Wagner:1988aa, Arribas:1993aa,Mediavilla:1995aa} and radial inflow due to the bar \citep{Schommer:1988aa} the primary choices. \citet{van-Gorkom:1983aa} and \citet{Schommer:1988aa} also found extended 6 and 20 cm radio emission spatially coincident with the ionization cones suggesting the presence of a jet. Further complicating interpretations was the finding of a strong isophotal twist ($\Delta$P.A. $\approx$ 60\degr{}) in the central $\approx5$\arcsec{} from NIR imaging that suggested the presence of a secondary nuclear bar. 

This secondary nuclear stellar bar was thought to potentially be the driving mechanism to bring gas from the circumnuclear ring to the SMBH. However, \citet{Petitpas:2002aa} found no significant nuclear molecular gas bar coincident with the stellar bar based on spatially resolved CO (1--0) data. Instead they note a double peaked morphology of the CO emission that straddles the centre but is perpendicular to the ionization cones and suggest that the NGC 5728 might have experienced a recent minor merger that could cause the off-center CO emission. 

All of these intriguing components contained with NGC 5728 make it an ideal system to study the fuelling and feedback of AGN. In this Paper, we report on a combined analysis of recent \textit{Atacama Large Millimeter/submillimeter Array} (ALMA), \textit{Very Large Telescope} (VLT) MUSE, and VLT/SINFONI observations that jointly probe the morphology and kinematics of the stars, cold molecular gas, hot molecular gas, and warm ionized gas on both large and small spatial scales. Recently, \citet{Durre:2018ab} and \citet{Durre:2018aa} studied NGC 5728 using both the MUSE and SINFONI datasets. Much of our conclusions and interpretation are consistent, but we will note  where we disagree and where their analysis enhances ours. Throughout this Paper, we assume a $\Lambda$CDM cosmology with $H_0 = 70$ \kms{} Mpc$^{-1}$ and $\Omega_{\rm M} = 0.3$. 

\section{Observations, Data Reduction, and Analysis Methods}

\subsection{SINFONI Observations}
NGC 5728 was observed by VLT/SINFONI \citep{Eisenhauer:2003eh, Bonnet:2004ab} as part of the \textit{Luminous Local AGN with Matched Analogues} (LLAMA) program \citep{Davies:2015uq}, a volume limited survey of nearby X-ray selected AGN with SINFONI and XSHOOTER. Within this program, all targets were observed by SINFONI using the H+K grating (1.4 -- 2.5 \micron) and AO mode providing integral field spectroscopy with spectral resolution $R\sim1500$ over a roughly 3\arcsec{}$\times$3\arcsec{} field of view (FOV). A standard Object-Sky-Object observation sequence was used with each integration lasting 300 seconds and the Object exposures dithered by 0.3\arcsec. Observing blocks for NGC 5728, under program ID 093.B-0057 (PI R. Davies), were carried out on the nights of February 23, March 06, May 09, June 04, and June 25, 2015 resulting in 25 Object integrations and 12 Sky integrations for a total on-source time of 125 minutes.  

The raw SINFONI data were reduced using the custom reduction package \textsc{spred} \citep{Abuter:2006aa} which performs all the standard reduction steps needed for near-infrared spectra as well as additional routines necessary to reconstruct the data cube. We further applied the routines \textsc{mxcor} and \textsc{skysub} \citep{Davies:2007aa} to improve the subtraction of OH sky emission lines and \textsc{lac3d}, a 3D Laplacian Edge Detection algorithm based on \textsc{lacosmic} \citep{vanDokkum:2001pg}, to identify and remove bad pixels and cosmic rays. Telluric correction and flux calibration were carried out using B-type stars. Finally, we also corrected for differential atmospheric refraction, a wavelength dependent spatial offset due to the atmosphere, using a custom procedure described in \citet{Lin:2018aa}.

The final SINFONI data cube covers a FOV of 3.5\arcsec{}$\times3$.7\arcsec{} with a pixel scale of 0.05\arcsec. With AO, we achieved a point spread function (PSF) full width at half maximum (FWHM) of 0.15\arcsec{} based on Gaussian fitting of the unresolved continuum nuclei from the AGN in the LLAMA sample. This translates to a physical spatial resolution of 28 pc over a physical FOV of 660$\times$700 pc. 

\subsection{MUSE Observations}
MUSE is a second generation VLT instrument consisting of 24 integral field units that span the wavelength range 4650--9300 \AA{} \citep{Bacon:2010aa}. We downloaded archival MUSE data of NGC 5728 from the MUSE-DEEP collection, an ESO Phase 3 data release that provides fully reduced, calibrated, and combined MUSE data for all targets with multiple observing blocks. The original observations were performed over two observing blocks on April 3 and June 3, 2016 for program ID 097.B-0640 (PI D. Gadotti), and were carried out in seeing limited Wide Field Mode that results in a 1\arcmin$\times$1\arcmin{} FOV sampled with a pixel scale of 0.2\arcsec. The total exposure time for the combined MUSE cube is 79 minutes and Gaussian fitting of a foreground star within the FOV show a PSF FWHM of 0.62\arcsec. The MUSE data cube thus covers an 11.3 kpc square region with a spatial resolution of 177 pc at the distance of NGC 5728. 

\begin{figure*}
\includegraphics[width=\textwidth]{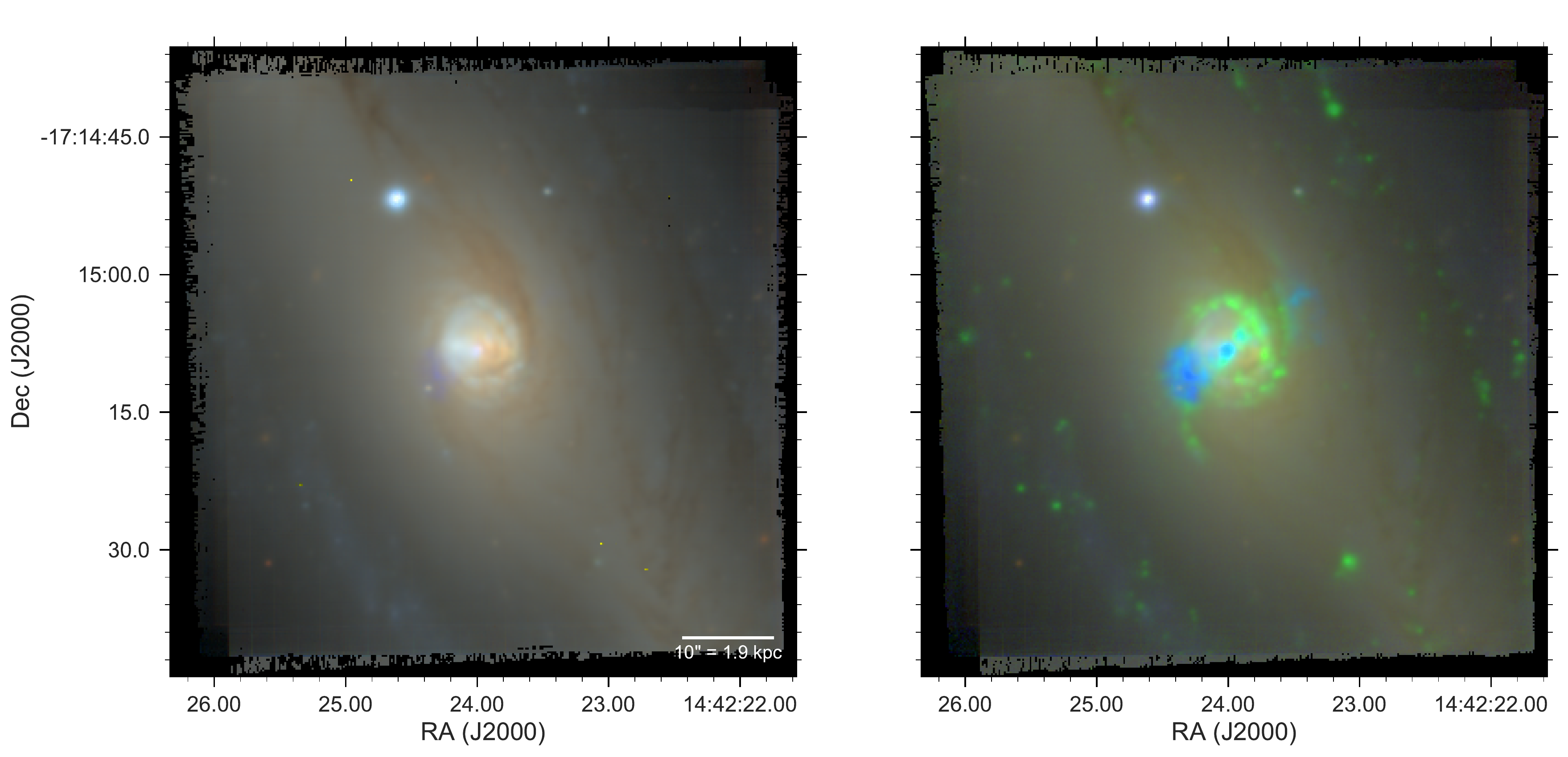}
\caption{\textit{Left panel:} Three color \textit{gri} image obtained by integrating the MUSE cube between wavelengths of 4800--5830 \AA{} for $g$-band, 5380--7230 \AA{} for $r$-band, and 6430--8630 \AA{} for $i$-band. \textit{Right panel:} Three color [OIII] (blue), \halpha{} (green), i-band (red) image using a narrow band filter to highlight [OIII] (5028--5095 \AA) and \halpha{} emission (6578--6664 \AA). \label{fig:muse_rgb}}
\end{figure*}

Figure~\ref{fig:muse_rgb} shows a three color \textit{gri} image of NGC 5728 from the MUSE cube. As seen in the HST imaging in Figure~\ref{fig:hst_rgb_fov}, the MUSE observation covers a large portion of the large-scale bar which contains long dust filaments and a blue circumnuclear disk/ring. In the right panel, we highlight strong [OIII] and \halpha{} emission by using a simple narrow band filter for the blue and green colours of the image while the red colour is from the i-band. 

\subsection{ALMA Observations}
ALMA observed NGC 5728 on May 15, 2016 as part of project 2015.1.00086.S (PI N. Nagar) for a total on-source time of 29 minutes in the C36-3 configuration resulting in an angular resolution of 0.56\arcsec$\times$0.49\arcsec (PA = -79.4\degr) in Band 6. This observation was part of an ALMA program to map CO (2--1) in the nuclear regions of 7 local AGN \citep{Ramakrishnan:2019aa}. The spectral setup was designed to center one of the spectral windows on the redshifted CO (2--1) emission line ($\nu_{\rm rest}=230.538$ GHz) for NGC 5728.

We used the Common Astronomy Software Applications package \citep[CASA;][]{McMullin:2007aa} to process the data. Within the \textsc{tclean} routine, we applied Briggs weighting with \textit{robust} = 0.5 to reproduce the data cube at 2.54 \kms{} velocity resolution and a peak residual (parameter \textit{threshold}) of 2.5 mJy. All resulting data cubes were primary beam corrected and a 1.3 mm continuum image was extracted by combining and integrating the remaining three spectral windows without the CO (2--1) line.

\subsection{HST Observations}
The final dataset we primarily use for this work is archival multiband HST imaging. HST observed NGC 5728 in the F336W, F438W, F814W and F160W bands with the \textit{Wide Field Camera 3} instrument as part of program 13755 (PI J.\ Greene). We downloaded the standard pipeline reduced images from the \textit{Mikulski Archive for Space Telescopes}\footnote{\url{https://mast.stsci.edu/portal/Mashup/Clients/Mast/Portal.html}}.  

\begin{figure*}
	\includegraphics[width=\textwidth]{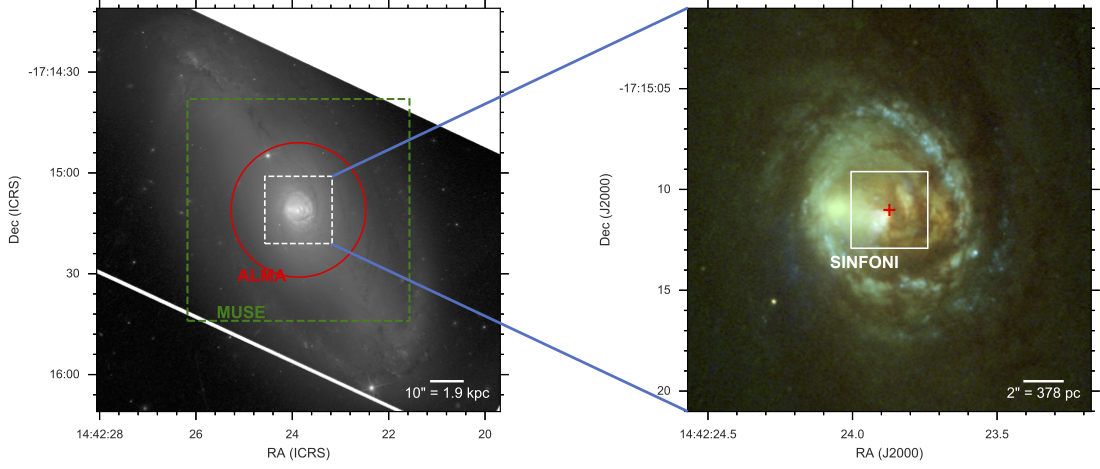}
    \caption{NGC 5728 as observed by \textit{HST}/WFC3. The left panel is a grayscale image of only the F814W observation to highlight the primary bar and bright circumnuclear region. The right panel is a three colour zoom-in of the region outlined by the white dotted box in the right panel. Red, green, and blue colours are from the F814W, F438W, and F336W filters. In the left panel, the green dashed line indicates the FOV of the MUSE observation while the red circle indicates the the FOV of the ALMA observation. The white box in the right panel indicates the FOV for the SINFONI observations and the red cross shows the position of the AGN from VLBI (see Section~\ref{sec:agn_pos}). In both panels, North is up and East is to the left.} 
    \label{fig:hst_rgb_fov}
\end{figure*}

Figure~\ref{fig:hst_rgb_fov} shows the F814W (\textit{left}) with a three colour zoom-in of the circumnuclear region that combined the F814W, F438W, and F336W images as red, green, and blue colours respectively. The larger scale F814W image also has overplotted the FOV's of the ALMA (red circle) and MUSE (green box) datasets. Both ALMA and MUSE cover a large portion of the primary bar and the entire circumnuclear region. The FOV of the SINFONI cube (white box) is overplotted on the three colour zoom-in that shows we are probing the very inner regions of NGC 5728. The combination of all of these datasets allow us to consistently trace gas and structure from several kpcs down to tens of pc as well as the interaction between the host galaxy and the central AGN.

\subsection{Astrometric Alignment and AGN Position}\label{sec:agn_pos}
To ensure all of the data sets are registered to a common coordinate system, we implemented the following procedure. Luckily at least one bright foreground star lies within the FOV of every HST image and the MUSE cube. Starting with the F336W, F438W, and F814W HST images, we matched the bright stars to GAIA DR2 sources \citep{Gaia-Collaboration:2018aa}. To make corrections, we simply updated the WCS parameters by applying small pixel shifts until the positions matched between the images and the GAIA DR2 values. These corrected HST images were then used to correct both the NICMOS F160W image and the MUSE cube, again by applying pixel shifts until the locations of the stars in common matched. 

For the SINFONI cube, no foreground stars appear in the FOV, therefore we first produced an H-band continuum image from the cube and used it to match distinct features between it and the HST F160W image. In particular, we aligned the brightest pixel in the SINFONI H-band continuum image with the brightest pixel in the HST F160 image and then checked to ensure other features such as the nuclear dust lane were aligned as well. This gives us an astrometric accuracy of $\approx0.25$\arcsec (i.e.\ the angular resolution of HST at 1.6 \micron) between SINFONI and the rest of the datasets. Finally, we did not apply any corrections to the ALMA data given the high astrometric accuracy inherent in radio interferometric observations. %We did however use the 1.3 mm continuum image to define the location of the AGN in NGC 5728. The bright point-like continuum source is likely emission from either cold dust associated with the parsec-scale molecular torus obscuring the AGN or compact synchrotron emission from the AGN itself. In either case, the continuum source provides a good reference for the central AGN given the spatial resolution of our datasets. The J2000 coordinates we derive from the continuum image are RA = 14:42:23.8737 and DEC = -17:15:11.012 in the ICRS coordinate system. Throughout the rest of this paper, when the ``AGN position'' is mentioned, these are the coordinates that are referenced.

For the position of the AGN, we used VLBI observations from The Megamaser Cosmology Project (C.~Y.~Kuo 2018, private communication). The coordinates of the AGN are RA = 14:42:23.872 and DEC = 17:15:11.016 with an accuracy of a few tens mas. The position is shown in the right panel of Figure~\ref{fig:hst_rgb_fov} as a red cross and will be used as the reference point for all following figures. This position matches very well both the peak of 1.3 mm continuum emission from ALMA (see Figure~\ref{fig:zoom_maps}) as well as the peak of the warm molecular gas traced by \warmH{} emission from SINFONI (see Figure~\ref{fig:sinfoni_fits}). 

\subsection{Spectral Fitting}

\subsubsection{Continuum Fitting and Subtraction}
The data reduction for the ALMA cube includes already a continuum subtraction however both the SINFONI and MUSE cube contain significant continuum signal. To remove the continuum as well as stellar absorption features, we use the \textit{Penalized Pixel-Fitting}\footnote{\url{http://www-astro.physics.ox.ac.uk/~mxc/software/\#ppxf}} (\textsc{pPXF}) routine \citep{Cappellari:2004aa,Cappellari:2017aa} together with the latest (v11.0) single stellar population (SSP) models based on the extended MILES stellar library \citep[E-MILES;][]{Vazdekis:2016aa}. We chose the models produced from the Padova 2000 stellar isochrones with a Chabrier initial mass function \citep{Chabrier:2003aa}. 
The SSP models range from stellar population ages of 63 Myr--14.13 Gyr spaced logarithmically and 6 metallicities between -1.71 and 0.22. 

For the MUSE cube, both to reduce the computation time and improve the signal-to-noise (S/N) ratio in the outer regions, we chose to utilize Voronoi binning before running \textsc{pPXF}. We used the method from \citet{Cappellari:2003aa} and the \textsc{voronoi\_2d\_binning}\footnote{\url{http://www-astro.physics.ox.ac.uk/~mxc/software/\#binning}} Python routine to bin the MUSE cube such that each bin contains a minimum S/N in the continuum of 20. %We decided against using Voronoi binning on the SINFONI cube to preserve the natural spatial resolution. 

\begin{figure*}
    \includegraphics[width=\textwidth]{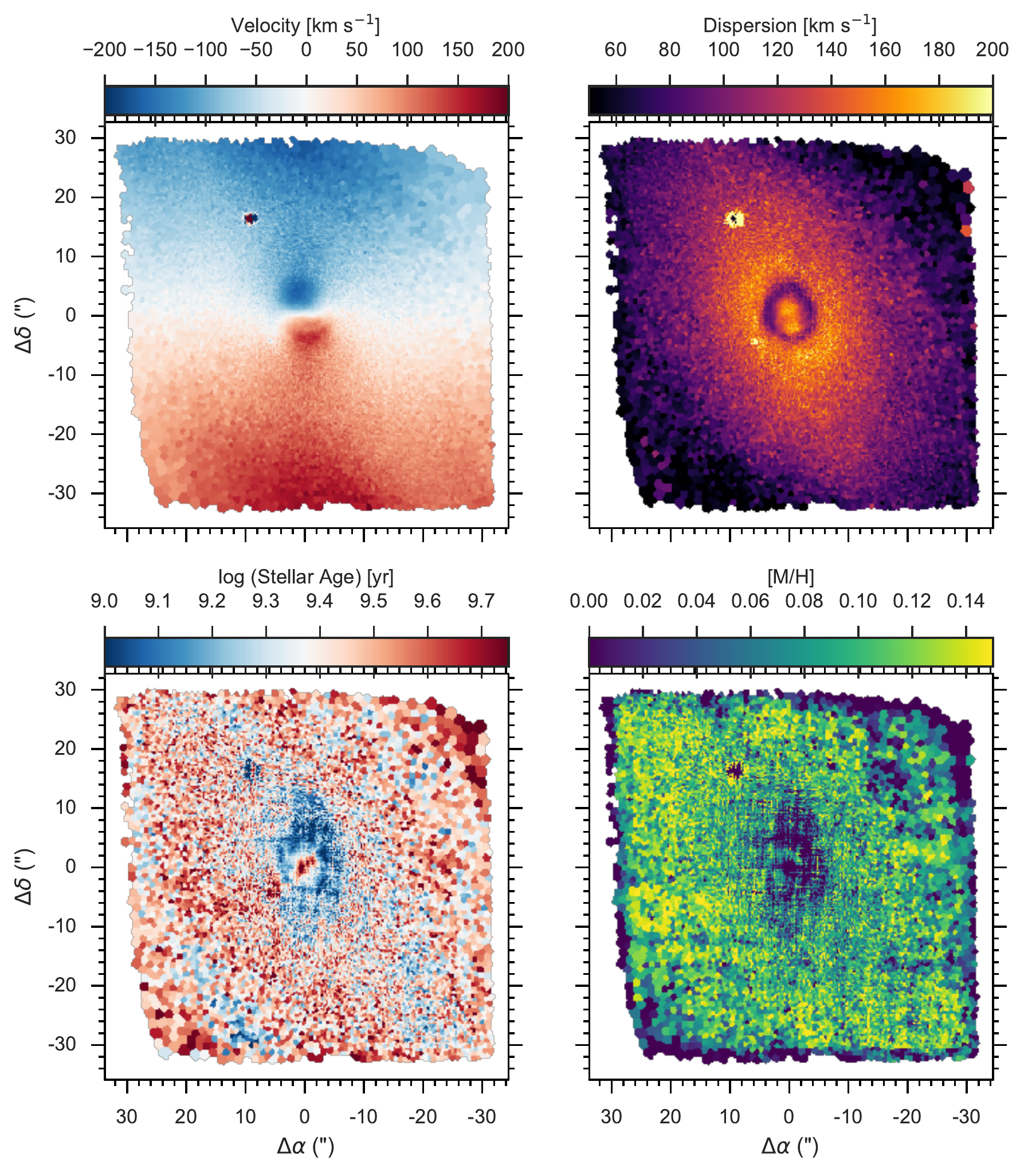}
    \caption{Results from fitting the MUSE cube with \textsc{ppxf}. The top row plots the stellar velocity (\textit{left}) and dispersion (\textit{right}). The bottom row plots the mean light-weighted stellar population age (\textit{left}) and mean light-weighted metallicity (\textit{right}). In all frames, North is up and East is to the left.
     \label{fig:stellar_fits}}
\end{figure*}

Each binned or single MUSE and SINFONI spaxel was fit with \textsc{ppxf}. For the MUSE spectra, we used the full range of metallicities and stellar population ages but we also incorporated regularization to better interpret the best fit metallicities and ages since the fitting of galaxy spectra is an ill-posed problem \citep{Cappellari:2017aa}. For the SINFONI spectra, because they only cover the near-infrared and are relatively insensitive to young stellar population ages, we only fit using models with ages of 0.1, 0.5, 1.0, 5.0, 10.0, and 14.1 Gyr. We also did not use regularization because our only concern for the SINFONI cube is the removal of the continuum and determination of the stellar kinematics. Because both the MUSE and SINFONI spectral resolutions are only moderate, we only fit for the stellar velocity and velocity dispersion. We also included an additive fourth degree polynomial in the fit to model any emission from the central AGN. To create the continuum subtracted MUSE cube, we renormalised the best fit binned spectra to the median continuum level of each individual spectrum before subtraction. Figure~\ref{fig:stellar_fits} shows the results of our \textsc{pPXF} fits for the MUSE cube.

\subsubsection{Emission Line Fitting}
To measure the bulk distribution and kinematics of the molecular and ionized gas in NGC 5728, we fit single Gaussians to the relevant emission line tracers of the gas. In particular, we fit the CO (2--1) line ($\nu_{\rm rest}=230.538$ GHz)  in the ALMA cube; the \hbeta, [OIII] doublet, [OI], [NII] doublet, \halpha, and [SII] doublet lines in the MUSE cube; and the [FeII]$\lambda1.64$ \micron, [SiVI], \warmH, and \bry{} lines in the SINFONI cube. This set of lines represent the strongest emission lines observed in nearby galaxies and AGN and provide diagnostic power into the state of gas in a range of phases from cold molecular gas (CO (2--1)) to warm molecular gas (\warmH) to hydrogen recombination lines (\hbeta, \halpha, \bry) and forbidden lines ([OIII], [OI], [NII], [SII], [FeII], [SiVI]) from ionized gas in a range of ionization levels. 

For each line, we cut out the relevant sections of the continuum subtracted spectra that contain the single line or multiple lines in the case of the [OIII] doublet, \halpha{} and [NII] blended region, and [SII] doublet. We fixed the velocity and velocity dispersion to be the same for all doublets and fixed the intensity ratio to the expected theoretical value of 3 for the [OIII] and [NII] doublets. For the fit of the [SiVI] line, we also included Gaussian components for the nearby H$_{2}$ (1--0) S(3) and Br$\delta$ lines. As an estimate of the local noise around each line, we measured the root-mean-square (RMS) of the line-free regions. Using the RMS, we produced 100 simulated spectra for every emission line by randomly perturbing the original spectra assuming a Gaussian distribution with a standard deviation equal to the RMS. Each of the simulated spectra were fit in the same way as the original spectra and uncertainties on the integrated flux, velocity, and velocity dispersion were calculated from the standard deviation of the best fit values from the simulated spectra.  Finally, all velocity dispersions derived from MUSE and SINFONI data were corrected for instrumental broadening. For lines from MUSE, we assumed an instrumental dispersion of 49 \kms{} for lines in the [OIII] region and 71 \kms{} for lines in the \halpha{} region based on the spectral resolution curve as a function of wavelength in the MUSE User Manual\footnote{\url{https://www.eso.org/sci/facilities/paranal/instruments/muse/inst.html}}. For SINFONI, we assumed an instrumental dispersion of 65 \kms{} based on the width of OH sky lines. Figures \ref{fig:muse_1_fits}, \ref{fig:muse_2_fits}, \ref{fig:sinfoni_fits}, and \ref{fig:alma_fit} show the results of our single Gaussian fitting for all of the emission lines studied in this paper over the full FOV for each instrument. Figure~\ref{fig:key_maps_zoom} shows the results for CO (2--1), \halpha, and [OIII] zoomed into the central 10\arcsec x 10\arcsec.

\begin{figure*}
	\includegraphics[width=\textwidth]{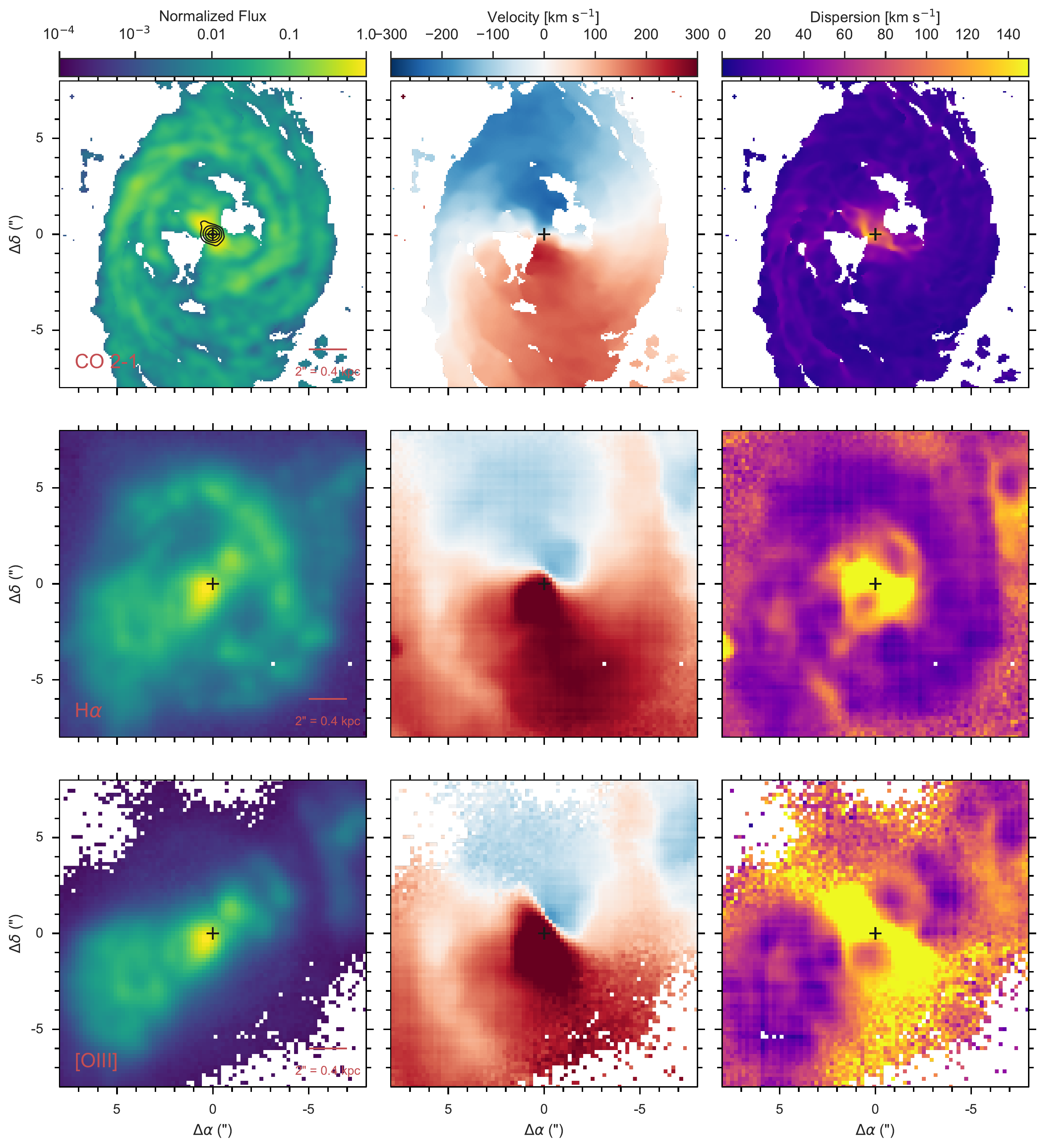}
	\caption{\label{fig:zoom_maps} Results of emission line fitting for the inner 8x8\arcsec{} region of NGC 5728 using a single Gaussian component. The top row shows the integrated flux (\textit{left}), velocity (\textit{middle}), and dispersion (\textit{right}) for CO (2--1) with the black contours plotting the 2$\sigma$, 5$\sigma$, 10$\sigma$, 20$\sigma$ contours of the 1.3 mm continuum. The middle and bottom rows show the flux, velocity, and dispersion for H$\alpha$ and [OIII] respectively. The black crosses in all panels show the location of the AGN from VLBI (see Section~\ref{sec:agn_pos}).}
\end{figure*}

\section{Results}

\subsection{Large Scales and the Primary Bar}

\begin{figure*}
	\includegraphics[width=\textwidth]{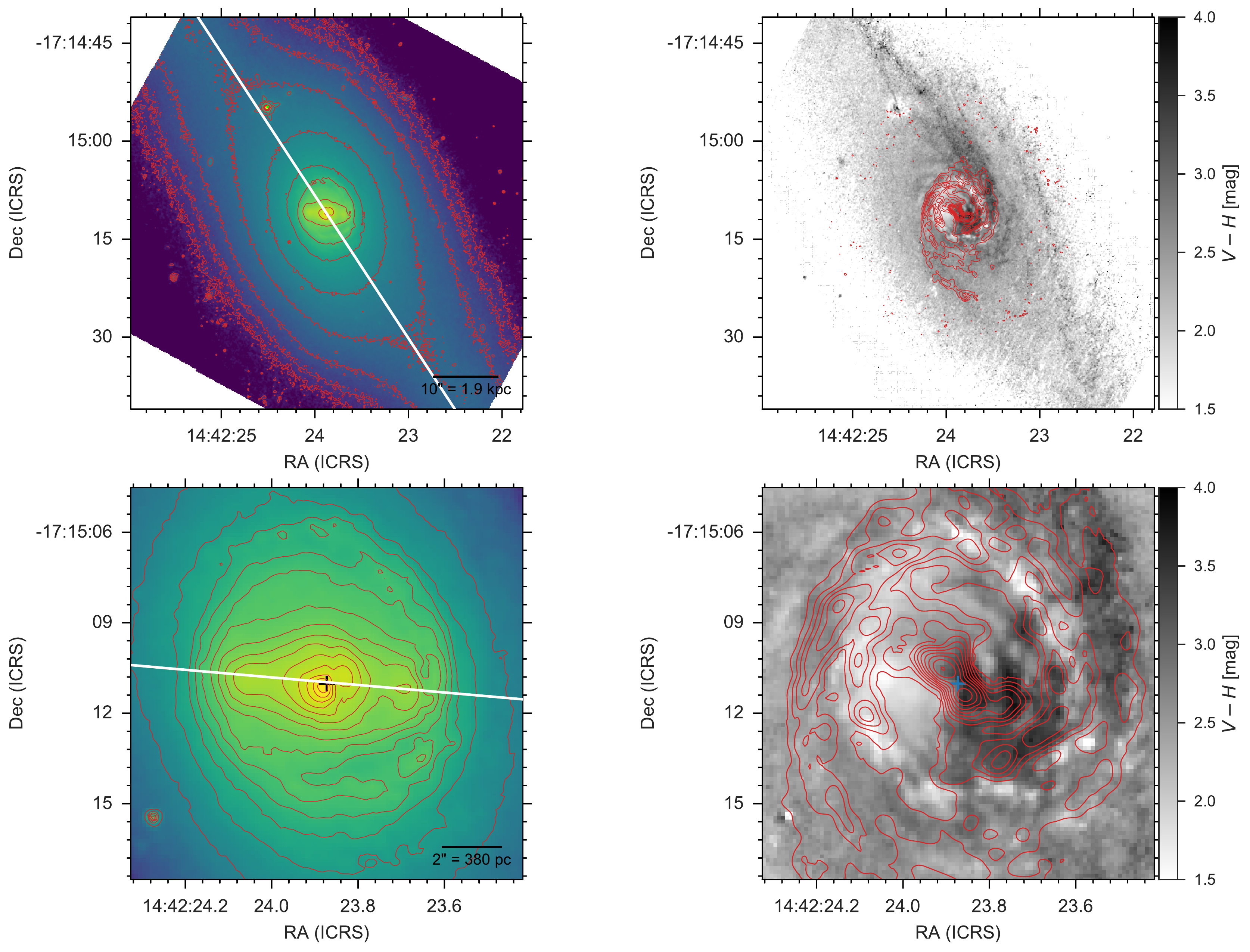}
	\caption{\label{fig:dustmaps} Large scale and circumnuclear zoom-ins of the H-band imaging and $V-H$ dust maps produced from F160W (H) and F438W (V) HST observations. The red contours in the left panels trace the continuum image to better outline the structure while the red contours in the right panels plot the CO (2--1) emission from ALMA. The white line in the top left shows the major axis of the large scale bar and in the bottom left the major axis of the nuclear stellar bar taken from the literature \citep{Schommer:1988aa,Prada:1999aa}}
\end{figure*}

We begin our investigation of NGC 5728 with the largest scales. As mentioned in Section~\ref{sec:ngc5728}, NGC 5728 is characterised on large scales by a primary bar surrounded by a ring of young stars. %Both the HST and MUSE data display these features as shown in Figures~\ref{fig:muse_rgb} and~\ref{fig:hst_rgb_fov}.
Continuum emission in general is elongated from the NE to the SW and the MUSE three-colour image (Figure~\ref{fig:hst_rgb_fov}) shows bluer emission encircling the inner, redder bar. These bluer emission regions in the large ring are also spatially coincident with clumps of \halpha{} emission indicating recent star formation. 

In the top left panel of Figure~\ref{fig:dustmaps}, we show in both colour and contours the HST F160W image which traces emission from older stars. Again, on the largest scales we observe primarily the large bar with the P.A. of the bar plotted as the white line, however towards smaller radii the stellar distribution becomes more and more spheroidal with the major axis shifting to more N--S. This is also the P.A. for the stellar kinematic major axis as shown in Figure~\ref{fig:stellar_fits} as well as the photometric major axis for the main disk that the bar lives within \citep{Schommer:1988aa}. \citet{Prada:1999aa} observed this isophotal twisting as well in large \textit{I}-band imaging and attributed it to the increasing influence of the central bulge. This would further make sense given the increasing stellar velocity dispersion which rises to $\sim170$ \kms{} before dropping at the circumnuclear ring (Figure~\ref{fig:stellar_fits}, top right panel) as well as having a relatively old mean age of $\sim5$ Gyr compared to both the larger main ring and circumnuclear ring which show ages of $\sim1$ Gyr (Figure~\ref{fig:stellar_fits}, bottom left panel).

Roughly straight dust lanes can also be seen in Figure~\ref{fig:muse_rgb} and~\ref{fig:hst_rgb_fov} running the length of the bar with the more prominent dust lane occurring in the northern half and indicating that the west side of the galaxy is the near side \citep{Schommer:1988aa}. To further enhance and highlight the dust lanes, we show in the top right panel of Figure~\ref{fig:dustmaps} a $V-H$ map constructed from the F438W and F160W HST images. Areas with high extinction would have large $V-H$ values and appear dark in the map. Indeed the northern dust lane reveals itself as a dark band extending from the NE to the circumnuclear region. We also plot on top of the $V-H$ map contours showing the location of CO (2--1) emission. Interestingly, the northern dust lane lacks strong CO emission and it is instead the southern dust lane that appears as a single spiral arm connecting to the circumnuclear ring. This morphology is consistent with the observations of CO (1--0) emission from \citet{Petitpas:2002aa} however they attribute the southern arm as potentially due to a larger ring structure or evidence of a past merger event. Instead it is quite clear within our datasets that it is associated with a fainter dust lane on the far side of the galaxy. It is unclear however why CO (2--1) is missing in the northern dust lane. It is possible that molecular gas here is more diffusely distributed over larger scales that were resolved out of the interferometric observations. 
 
As hydrodynamic simulations show \citep[e.g.][]{Athanassoula:1992aa,Maciejewski:2002aa}, two straight dust lanes are expected to appear along the bar major axis due to the formation of principal shocks which compress the gas and dust. This occurs due to the gradual shifting of the gas orbits from the $x_1$ family (i.e. parallel to the bar major axis) to the $x_2$ family (i.e. perpendicular to the bar major axis). As gas crosses these shocks, it loses angular momentum resulting in inflow towards the centre \citep[e.g.][]{Athanassoula:1992aa,Maciejewski:2002aa}. As we will show in the next section, we do observe kinematic signatures of inflow at the locations where the dust lanes connect to the central circumnuclear ring strongly suggesting that gas is being driven from the outer disk to the circumnuclear region via the bar. Assuming the gas in the bar is rotation dominated it is expected to settle into a ring \citep[e.g][]{Piner:1995aa} or nuclear spiral \citep{Maciejewski:2002aa,Maciejewski:2004aa}.

\subsection{The Circumnuclear Ring}

As many others have before we observe a central circumnuclear ring within the central $\approx1.3$ kpc in projected radius. Multiple tracers highlight this structure through both their emission and kinematics.

Both the HST and MUSE continuum imaging show a distinct ring of clumpy, blue emission indicative of young hot stars. This is confirmed in the bottom left panel of Figure~\ref{fig:stellar_fits} which shows the mean light-weighted stellar age. The circumnuclear ring outlined by blue continuum emission also has a distinctly younger mean light-weighted stellar age (1 Gyr vs 5 Gyr) compared to the surrounding region. This fits with the LLAMA finding that on average AGN show a significant young ($<30$ Myr) stellar population within the central few hundred parsecs compared to inactive galaxies based on stellar population modelling of high spectral resolution VLT/X-Shooter spectra (Burtscher et al. in preparation). Further, \citet{Riffel:2009cm} found a significant population of intermediate age stars in stellar population modelling of NGC 5728 using a NIR  IRTF spectrum. The stars in the ring further show evidence for lower metallicity which could be an indication that the increased nuclear activity is due to a recent minor merger.

The circumnuclear ring is further revealed kinematically in the stars via strong circular rotation and a drop in velocity dispersion (Figure~\ref{fig:stellar_fits}, top row). While on large scales the stellar velocity peaks at the edge of the FOV and with a P.A.$\sim$0\degr, there is a clear kinematically decoupled component with velocity peaks at $\sim$4\arcsec{} (760 pc) radius and a P.A.$\sim10$\degr. The dispersion map shows the ring as a ring of decreased velocity dispersions compared to the surrounding regions (70 \kms{} vs. 170 \kms{}) which spatially corresponds to the ring of younger mean stellar ages. The reduced velocity dispersions can be explained as the primary stellar population not yet thermalising out of their birth clouds and still moving with regular rotation. These so-called dispersion ``drops'' have routinely been used in long slit observations of nearby galaxies as markers of nuclear star formation \citep[e.g.][]{Emsellem:2001aa} . 

\begin{figure*}
	\includegraphics[width=\textwidth]{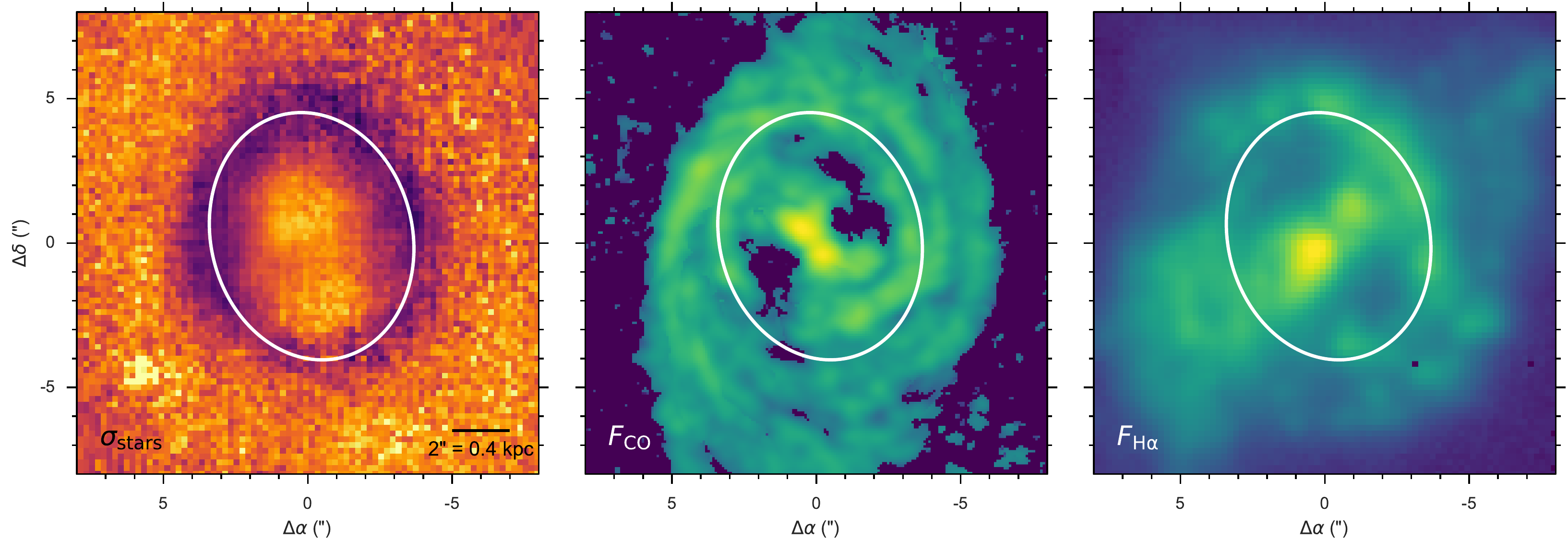}
	\caption{Comparison between the stellar velocity dispersion (\textit{left}), CO (2--1) flux (\textit{middle}), and \halpha{} flux (\textit{right}). The white ellipse plots the location of the lowered stellar velocity dispersion in each of the panels to the show the spatial correspondence between the decreased stellar dispersion and the rings of molecular gas and ongoing star formation. Colours in each panel are the same as what is shown in Figures~\ref{fig:stellar_fits} and \ref{fig:zoom_maps}.}
	\label{fig:circum_ring}
\end{figure*}

All of the stellar tracers suggest recent star formation so there should be large amounts of both molecular gas from which the stars were born out of and ionized gas from photoionization by the young stars. Indeed, Figure~\ref{fig:circum_ring} shows that exactly where we find the ring of lowered velocity dispersions, we also observe a strong ring of CO (2--1) emission as well as a ring of \halpha{} clumps outlining the star forming HII regions.

The white ellipse plotted in Figure~\ref{fig:circum_ring} was chosen to directly follow the ring of lowered stellar velocity dispersion. The parameters of the ellipse have a P.A. of 14\degr{}, a semimajor axis 822 pc and an axis ratio ($b/a$) of 0.8. If we make the assumption that the ring is intrinsically circular then the inclination is related to the axis ratio through $\cos(i) = b/a$. This leads to an estimate of the inclination of $\approx37$\degr{} which is more face-on compared to the large scale photometrically determined inclination of 48--55\degr{} \citep{Schommer:1988aa,Davies:2015uq}. This could be due to our assumption of circularity, since its possible the stars in the ring are on elliptical $x2$ orbits associated with the large scale bar.  These rings of young nuclear star formation associated with low velocity dispersions seem to be a common characteristic of local Seyfert galaxies as many studies using IFU data have found them within the central kpc \citep[e.g.][]{Riffel:2010aa,Riffel:2011ac,Diniz:2017aa,Dametto:2019aa}

\subsubsection{Gas Excitation}
The MUSE and SINFONI IFU cubes allow for us to spatially resolve the distribution of emission line ratios and assess the dominant mechanism for gas excitation. Gas excitation mechanisms can be identified using popular line ratio diagrams with specific thresholds or demarcations that separate between the different mechanisms which include photoionization by a central AGN, ionization by UV photons from HII regions, and low-ionization nuclear emission line regions (LINER). %Traditionally, these line ratio diagrams were constructed using a large sample of galaxy integrated spectra or single fibre nuclear spectra (i.e. SDSS), but with the advent of IFUs, we now have the ability to spatially resolve these ratios and asses the relative importance of each mechanism throughout a galaxy.

 \begin{figure*}
	\includegraphics[width=\textwidth]{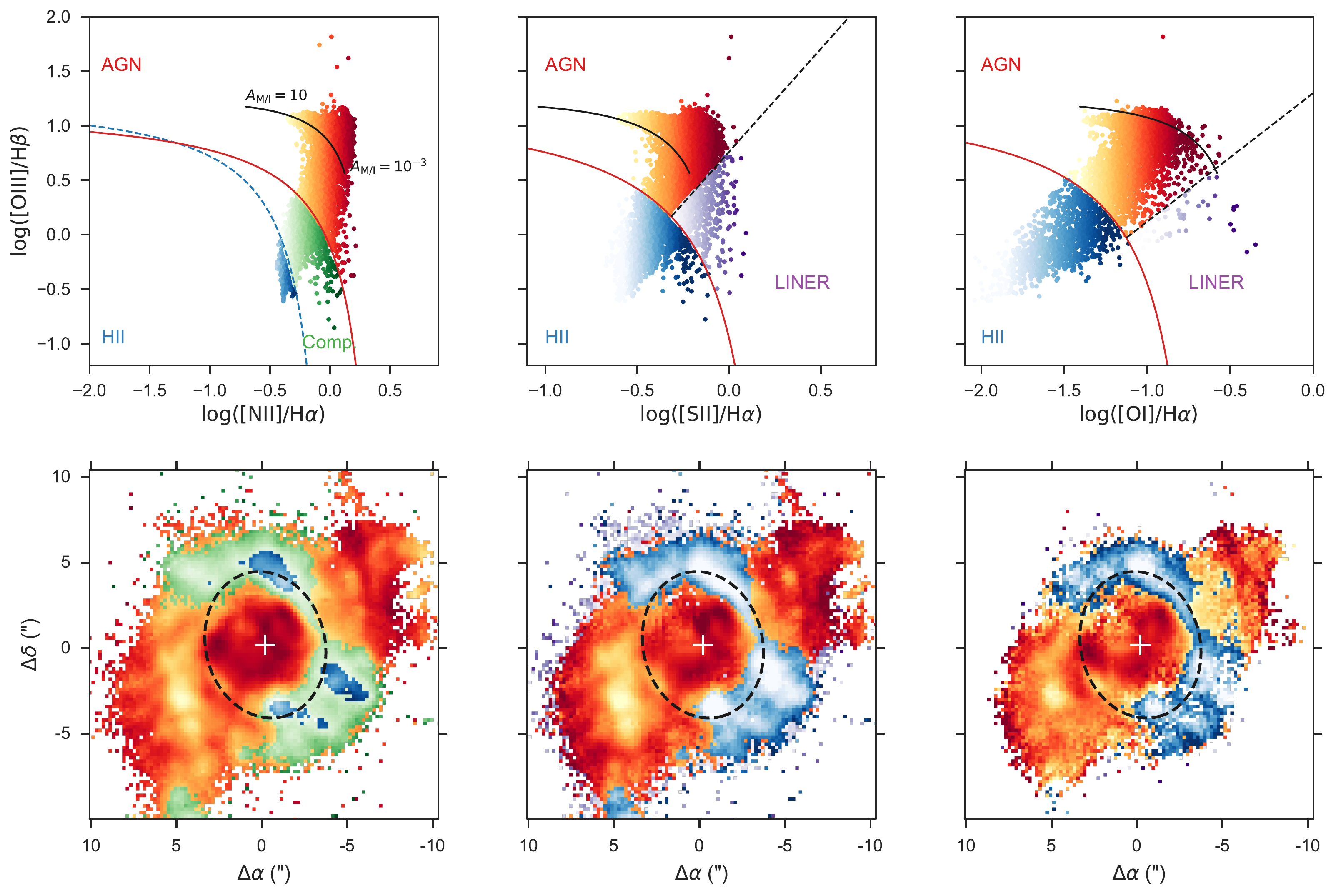}
	\caption{\label{fig:bpt_diagrams} Spatially BPT maps of NGC 5728 derived from emission line fitting from the MUSE cube. The top row plots the BPT line ratio diagrams with each dot representing a single spaxel. To be included in the diagram, the corresponding emission lines had to be detected with S/N $>$ 3. Points are colour-coded according to their location with respect to the  \citet[][blue dashed]{Kauffmann:2003mw}, \citet[][red solid]{Kewley:2001fj}, and \citet[][black dashed]{Kewley:2006uf} classification lines. The black solid lines plot the expected line ratios from \citet{Binette:1996aa} for an increasing ratio of the solid angle subtended by matter bounded and ionization bounded clouds ($A_{\rm M/I}$). The panels in the bottom row show the spatial location of each of the points coloured by their location in the corresponding top row panels. The black dashed ellipses outline the location of the circumnuclear ring seen in Figure~\ref{fig:circum_ring}. }
\end{figure*}

Figure~\ref{fig:bpt_diagrams} plots the three standard excitation diagrams \citep{Veilleux:1987vn,Baldwin:1981yq} for individual spaxels in NGC 5728. %These diagrams use the [OIII]/\hbeta{} line ratio together with the [NII]/\halpha{}, [SII]/\halpha{}, and [OI]/\halpha{} line ratios to distinguish between regions dominated by AGN, star-forming (SF), composite (SF+AGN or shocks), and LINER emission. The top row shows the line ratios plotted against each other with the blue dashed line in the [OIII]/\hbeta{} vs [NII]/\halpha{} indicating the empirical relation from \citet{Kauffmann:2003mw} and dividing pure star-forming galaxies from composite and AGN dominated galaxies. The red lines in all three diagrams in the top row show the ``maximal starburst'' lines calculated in \citet{Kewley:2001fj} and indicate the theoretical upper limits on the line ratios expected from models with only stellar photoionization. The black dashed lines in the [OIII]/\hbeta{} vs [SII]/\halpha{} and [OIII]/\hbeta{} vs [OI]/\halpha{} diagrams further split galaxies into Seyferts and LINERs and were determined empirically in \citet{Kewley:2006uf}. Where traditionally, single points in these diagrams indicated a single galaxy, each point in the top row of Figure~\ref{fig:bpt_diagrams} shows a single spaxel from the MUSE cube. The bottom row of Figure~\ref{fig:bpt_diagrams} shows the map of spaxels colour-coded by their location on the respective BPT diagram with blue, green, red, and purple representing SF, Composite, Seyfert, and the LINER regions respectively.
The circumnuclear region of NGC 5728 is dominated by gas photoionized by the AGN, as indicated by the yellow to red colours. The AGN-dominated region further follows the bicone shape of the strong [OIII] emitting region. HII regions instead follow a clear ring-like structure, encircling the bicone and tracing the ring-shaped emission seen in \halpha{}, the three colour HST image, and the molecular gas distribution that is indicated by the white ellipse in the bottom row Figure~\ref{fig:bpt_diagrams}. All of this definitively points to strong star formation occurring within the circumnuclear ring. %The continuous connection of the star-formation dominated region in the NW and lack of connection in the SE reinforces the interpretation that the bicone is spatially in front of the nuclear disk/ring to the SE and behind it to the NE. 

One interesting feature seen in the excitation diagram is a horizontal ``spike'' at high [OIII]/\hbeta{} ratios and pointing towards lower [NII]/\halpha, [SII]/\halpha, and [OI]/\halpha{} ratios. Within each of the diagnostic regions, we shaded the spaxels from light to dark based on the [NII]/\halpha{}, [SII]/\halpha{}, or [OI]/\halpha{} line ratio value. In the maps we can then see that the horizontal ``spike,'' indicated by yellow colours, is primarily occurring in a narrow region on the SE side of the circumnuclear ring and coinciding with where the ring intersects the AGN bicone. 

This feature has been seen recently in other nearby AGN with spatially resolved line ratio maps. \citet{Mingozzi:2018aa}, using MUSE data for the MAGNUM survey, explained these line ratios as tracing the ratio of the solid angle subtended by matter bounded and ionization bounded gas clouds ($A_{\rm M/I}$) as first put forth in \citet{Binette:1996aa}. In short, matter bounded clouds are optically thin throughout the entire cloud to the ionizing radiation field while ionization bounded clouds are optically thick and absorb all of the incident ionizing radiation. If matter bounded clouds dominate the solid angle we would expect more emission from high ionization lines compared to low ionization ones and vice versa if ionization bounded clouds dominate. 

\citet{Binette:1996aa} using photoionization modelling produced predictions for the line ratios expected given an $A_{\rm M/I}$ with the following relation:

\begin{equation}\label{eq:binette}
R^{i}(A_{\rm M/I}) = \frac{R_{\rm IB}^{i} + 0.568A_{\rm M/I}R_{\rm MB}^{i}}{1 + 0.568A_{\rm M/I}}
\end{equation}

\noindent where $R^{i}$ indicates a line ratio for a specific ion relative to \hbeta{} and $R_{\rm IB}^{i}$ and $R_{\rm MB}^{i}$ are constant ratios in the limiting cases for fully ionisation bounded clouds and matter bounded clouds respectively. We plot this sequence in the top panels of Figure~\ref{fig:bpt_diagrams} shown as a black line. Especially for the [NII]/\halpha{} plot, the increase of $A_{\rm M/I}$ perfectly explains the spike in the excitation diagram.

% For the spaxels located in the spike, we plot them in the bottom panels with an orange colour to show where in the galaxy they occur. The high $A_{\rm M/I}$ spaxels are all co-located in a region to the SE about 5\arcsec away from the AGN, well within the NLR similar to what was seen for the AGN in \citet{Mingozzi:2018aa}. However, as we showed previously, this region is well outside where we observe signatures of the outflow whereas \citet{Mingozzi:2018aa} found these regions to be associated with the outflow. 

However, we can also explain this feature by an increase in the ionization parameter. \citep[][; BN19]{Baron:2019aa} developed a simple method for measuring the ionization parameter using the [OIII]/\hbeta{} and [NII]/\halpha{} line ratios (see Equation 2 of BN19). The main assumption is only that the gas is photoionized by an AGN. Therefore, we calculated the ionization parameter, $U$, as a function of radius towards the SW using a series of annular sections described Sec. \ref{sec:electron_density} and shown in Fig.~\ref{fig:muse_annuli}.

\begin{figure}
\includegraphics[width=\columnwidth]{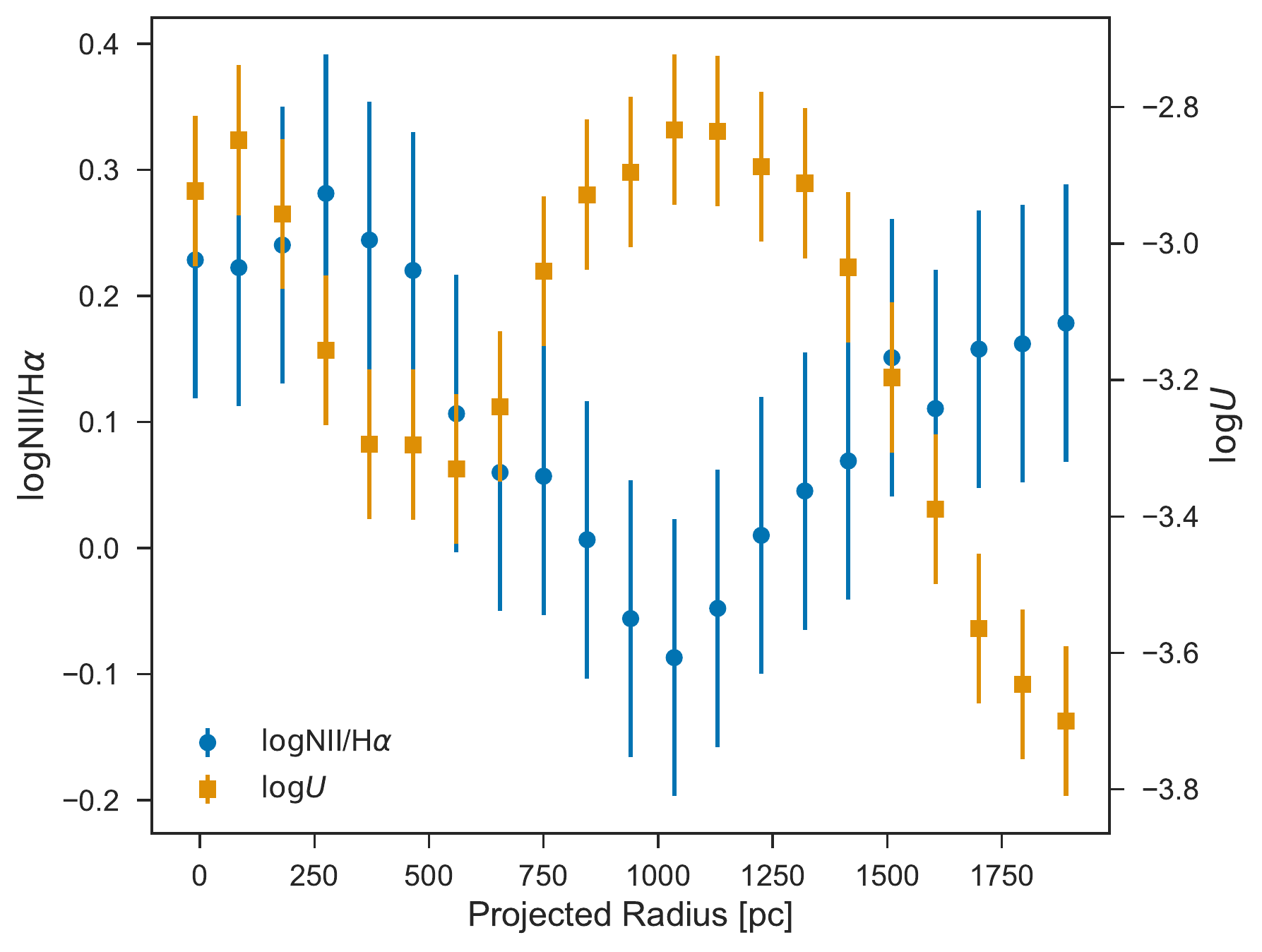}
\caption{\label{fig:logU} Radial profile towards the SW along the bicone of both the [NII]/\halpha{} line ratio and the ionization parameter, $U$. Radii have not been corrected for inclination and are simply the projected radius from the AGN.}
\end{figure}

Fig.~\ref{fig:logU} shows the radial profile along the bicone of both [NII]/\halpha{} and $U$. Indeed, exactly where we observe a decrease of [NII]/\halpha{}, the ionization parameter significantly increases. While both a change in $A_{\rm M/I}$ and a change in $U$ can explain the horizontal "spike" in the line ratio diagrams, we choose the simpler increase in $U$ model as our explanation. This would be consistent with an interpretation that this region is where the NLR bicone is intersecting the lower density circumnuclear ring since $U$ and gas density are inversely related.

%Instead for NGC 5728, based on the kinematics, it is likely gas still in the circumnuclear disk. The high $A_{\rm M/I}$ region seems to line up well with the HII and Composite regions that make up the star-forming ring. Therefore, our interpretation is that this region is where the NLR bicone is intersecting the circumnuclear disk. It is here where the gas in the disk has a direct view of the hard ionizing radiation from the AGN with little shielding, and creating a large ratio of high excitation matter bounded clouds which increases $A_{\rm M/I}$.

\subsubsection{Molecular Gas Mass}
We estimate the total molecular gas mass in the circumnuclear region from the summed CO (2--1) spectrum resulting from integrating the ALMA cube within a circular aperture with a radius of 6.5\arcsec{} (1.24 kpc). Because of the complex line profile, we simply integrated the spectrum between rest-frame velocities of $\pm375$ \kms{} to obtain a total CO (2--1) intensity ($S_{\rm C O}$) of 93.6 Jy \kms. The CO luminosity is calculated following \citet{Solomon:2005df}:

\begin{equation}\label{eq:lco}
L^\prime_{\rm CO} = 3.25\times10^{7}\,\frac{S_{\scriptscriptstyle \rm CO}R_{\scriptscriptstyle 12}D_{\scriptscriptstyle L}^{2}}{(1+z)\nu_{\scriptscriptstyle \rm rest}^{2}}\,\rm{K\,km\,s^{-2}\,pc^{2}}
\end{equation}

\noindent where $R_{12}$ is the conversion from CO (2--1) to CO (1--0) intensity, $D_{L}$ is the luminosity distance in Mpc, and $\nu_{rest}$ is the rest frequency of the CO (2--1) emission line in GHz. In this work we use $R_{12} = 1.4$, the value found for nearby star forming galaxies \citep{Sandstrom:2013en}.

To calculate a molecular gas mass, we must assume a value for $\alpha_{\scriptscriptstyle \rm CO}$, the conversion factor from $L^\prime_{\rm CO}$ to $M_{H_{2}}$. For this work, we use $\alpha_{\scriptscriptstyle \rm CO} = 1.1$ M$_{\sun}$ pc$^{-2}$/(K km s$^{-2}$). This is the value found for the central regions of nearby galaxies, likely due to the increased pressure and/or turbulence \citep{Sandstrom:2013en}. Using this, we find a total molecular gas mass within the central $\sim$1 kpc of NGC 5728 to be 1.3$\times10^{8}$ M$_{\sun}$. 

This total mass matches well with that found in \citet{Rosario:2018aa} who found a molecular gas mass of 10$^{8.53}$ M$_{\sun}$. Our value is 40\% of theirs, however the \citet{Rosario:2018aa} measurement is based on an JCMT observation with a beam size of 20.4\arcsec compared to the 6.5\arcsec aperture used here. This seems to be an indication that about 60\% of the total molecular gas is on scales larger than 1 kpc given the largest angular scale of the ALMA observation was 5.34\arcsec. We tested whether convolving the ALMA cube to the JCMT beam and integrating over a larger aperture increased the measured molecular mass but did not find any change, an indication that diffuse flux was resolved out due to the lack of short spacings during the ALMA observation.

\subsubsection{Disk Modelling}\label{sec:disk_modeling}
Kinematically, all three tracers (stars, molecular gas, and ionized gas) show strong signs of circular rotation (see Figures~\ref{fig:stellar_fits} and \ref{fig:zoom_maps}. Therefore, we can fit the velocity fields with a model of rotation to better constrain the inclination and dynamical mass driving the rotation.

\begin{figure*}
    \includegraphics[width=\textwidth]{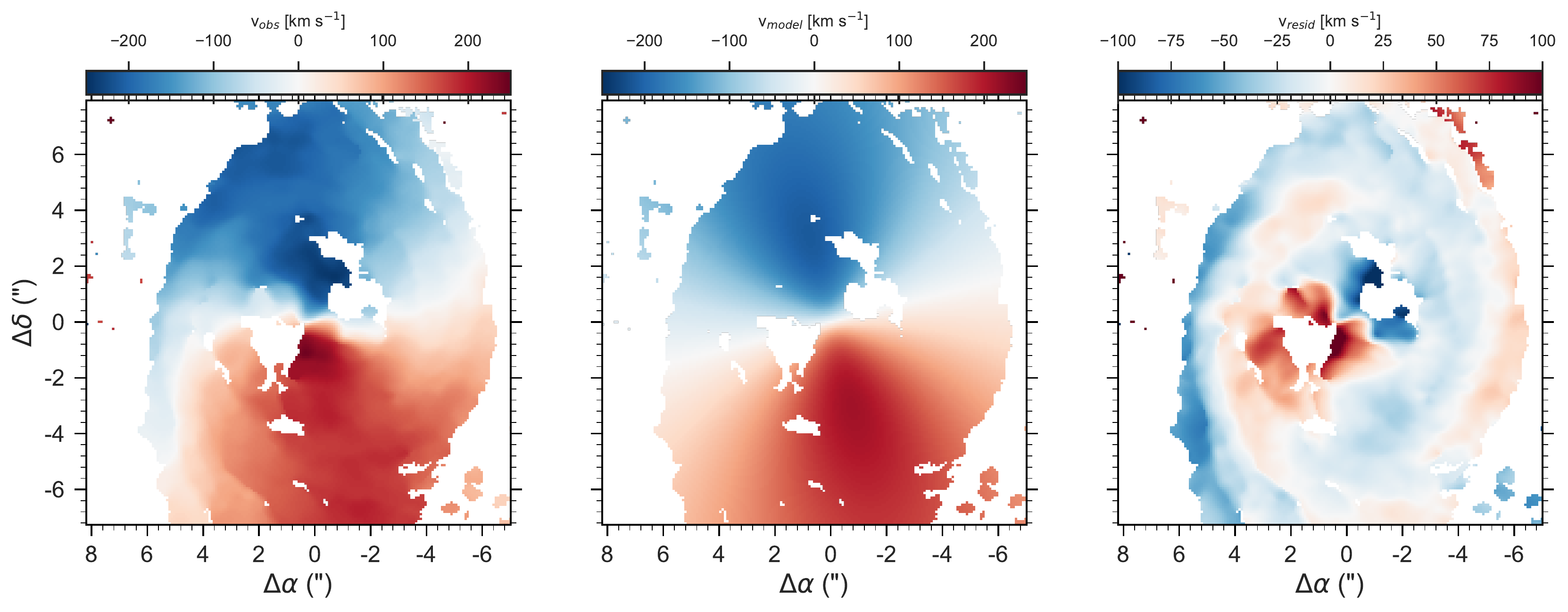}
    \includegraphics[width=\textwidth]{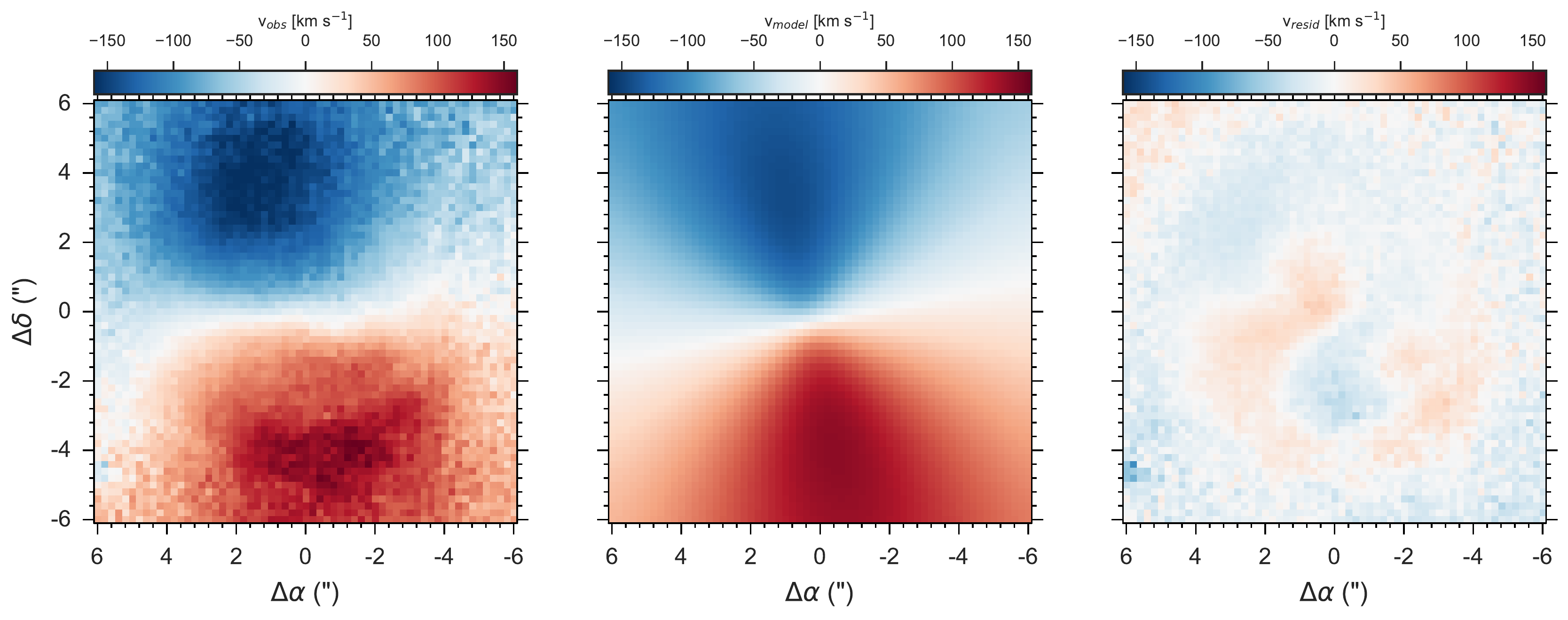}
    \caption{\label{fig:disk_model}Results of fitting the CO (2--1) (\textit{top row}) and stellar velocity field (\textit{bottom row}) with a rotating disk model. The observed velocity field is shown in the left panels, the best fit model in the middle panels, and the residuals in the right panels.}
\end{figure*}

We first fit both the CO (2--1) and stellar velocity fields using an updated Python version of \textsc{dysmal} \citep{Davies:2011aa}, now called \textsc{dysmalpy}, an MPE developed software package for modelling the dynamics of galaxies. \textsc{dysmalpy} works by first simulating the mass distribution of a galaxy and populating a three dimensional cube with the corresponding physical velocities. After rotating and inclining the galaxy to match the preferred line of sight, the physical 3D cube is integrated along the line of sight to produce an observed ``IFU'' cube which is finally convolved with both a PSF and line spread function. In this way, we account for all instrumental effects including beam smearing. These simulated cubes can then be used to compare and fit against observed data using either standard non-linear least squares or a Monte Carlo Markov Chain (MCMC) algorithm with \textsc{emcee} \citep{Foreman-Mackey:2013lr}.

We assumed a thin exponential disk mass distribution parameterised by the total dynamical mass ($M_{\rm dyn}$) and the effective radius ($r_{\rm eff}$) within which the enclosed mass is one half of $M_{\rm dyn}$. The inclination of the disk ($i_{\rm disk}$) and the position angle (P.A.$_{\rm disk}$) were also left as free parameters in the fitting. For the CO (2--1) model we fixed the centre by eye because we also masked out the inner 2.5\arcsec due to the presence of the non-circular motion described below. For the stellar velocity field we allowed the centre to vary between +/- 5 pixels in both the x and y direction.  We ran MCMC with 100 walkers for 100 burn-in steps and 400 sampling steps to determine the best fit parameters and uncertainties from the sampled posterior distribution using the median and the 16th and 84th percentiles.

Figure~\ref{fig:disk_model} shows the results of our fitting. For the molecular gas, we find a best fitting $\log M_{\rm dyn} = 10.372^{+0.002}_{-0.001}$ M$_{\sun}$, $r_{\rm eff} = 537^{+2}_{-2}$ pc, $i_{\rm disk} = 43.3\degr^{+0.1}_{-0.1}$, and P.A.$_{\rm disk} = 14.01\degr^{+0.04}_{-0.03}$. We note that the very small errors on the model parameters only incorporate the statistical uncertainties and thus are more reflective of the high S/N and high spectral resolution of the ALMA data. These results are also consistent with those found using the same data in \citet{Ramakrishnan:2019aa}. For the stellar velocity field, we find a best fitting $\log M_{\rm dyn} = 10.14^{+0.01}_{-0.01}$ M$_{\sun}$, $r_{\rm eff} = 640^{+10}_{-10}$ pc, $i_{\rm disk} = 42.1\degr^{+0.4}_{-0.5}$, and P.A.$_{\rm disk} = 10.9\degr^{+0.1}_{-0.1}$.

%Interestingly, the largest deviations from the model occur around the edges of the holes of molecular gas seen inside the ring. As shown later, the ionized gas that fills these holes show signs of an outflow, so it is possible that the enhanced velocities we observe here are due to outflowing molecular gas. The other locations we observe deviations from circular motion are in the outskirts of the disk. This is likely due to gas streaming radially along the large scale bar and entering the nuclear disk.

The residuals in the stellar velocity field all have an absolute value less than 40 \kms{} and the discrepancies are likely due to deviations from pure circular rotation since we expect stars in the circumnuclear regions to follow more elliptical $x1$ and $x2$ orbits. This can also be seen in measured velocity field where the line of nodes are not exactly perpendicular to the zero velocity line. These deviations are likely the reason for the discrepancy in the best fit model parameters between the molecular gas and stars. However, the good agreement between the models and the data as well as between the cold molecular gas and the stars indicates that our interpretation of a rotating circumnuclear ring is correct and that the young stars have been born out of the molecular gas and still exhibit the molecular gas kinematics. Finally we do note that the zero velocity line of both the stars and molecular gas are S-shaped which could be an indication of inflow of both stars and gas.

\subsubsection{Molecular gas outflow and inflow?}

\begin{figure}
	\includegraphics[width=\columnwidth]{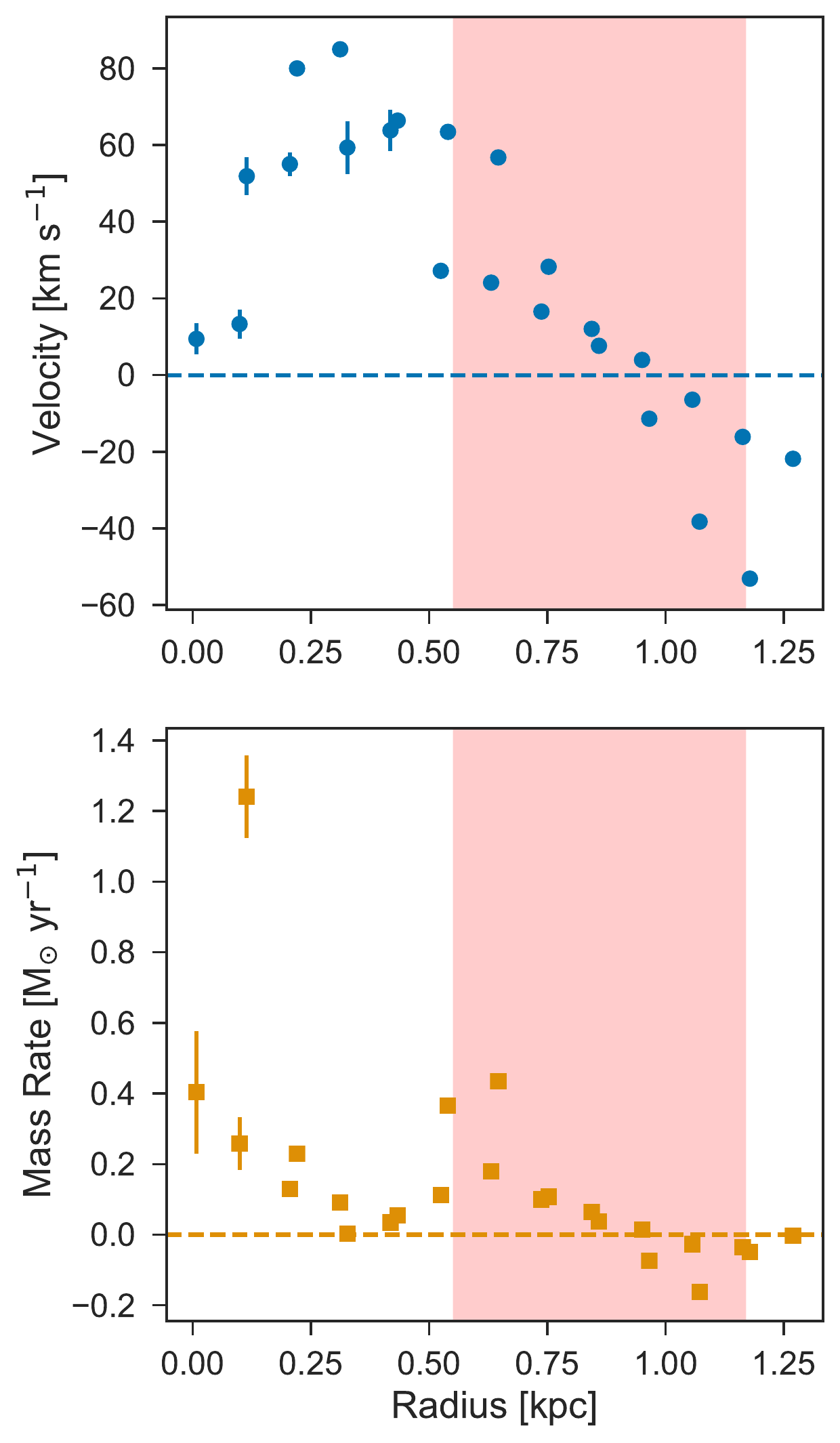}
	\caption{\label{fig:mol_inflow} Radial profile of the residual flow velocity (blue) and flow mass (orange) calculated from concentric annuli placed on the residual velocity field of the molecular gas after removing the best fit rotating disk model (See Figure~\ref{fig:disk_model}). Negative values correspond to outflow while positive values correspond to inflow. The dashed lines indicate the transition from outflow to inflow for the velocity and mass rate. The red shaded region shows the radii where we observe the molecular gas ring. Inner to the ring we potentially observe outflow while outside the ring we observe inflow.}
\end{figure}

The velocity residuals after removing circular rotation could be due to non-circular inflow and outflow. To investigate this, we placed circular apertures onto the residual velocity field spaced in increments of 0.56\arcsec{} along the minor axis of the best fit disk model. Placement along the minor axis ensures the velocity residuals are primarily due to radial motion rather than differences in tangential motion compared to circular rotation.

Radial inflow will appear as redshifted residuals on the near-side of the disk and blueshifted residuals on the far-side. We know based on the large scale dust maps and \halpha{} and [OIII] flux maps as well as the excitation of the ionized gas that the near-side is West of the kinematic major axis. Radial outflow motion will have the opposite sense as radial inflow. We calculate the total mass in the aperture by summing the integrated flux of all pixels in the aperture and using Equation~\ref{eq:lco} and $\alpha_{\rm CO}$ as before. The mass rate through each aperture is then $Mv/\delta r$ where $\delta r$ is the diameter of the apertures.

Figure~\ref{fig:mol_inflow} plots both the radial velocity estimates (top) and the mass rate (bottom). Positive values of each correspond to outflow while negative values correspond to inflow. It is from this plot that the residuals suggest an outflow velocity of $\sim$80 \kms{} and mass outflow rate of $\sim0.5$ $M_{\sun}$ yr$^{-1}$ within 1 kpc. We compare these values later to those found for the ionized gas.

Figure~\ref{fig:mol_inflow} also suggests radial inflow outside of 1kpc with velocities reaching 50 \kms. The mass inflow rate into the circumnuclear ring however is very small, only  $\sim0.1$ $M_{\sun}$ yr$^{-1}$. The strongest inflow residuals though do occur in the NW and SE edges of the ring, exactly where the dust lanes meet the ring and where we would expect to find signatures of inflowing gas.

%We can compare the mass inflow rate to the current mass accretion rate onto the SMBH using the following equation:
%
%\begin{equation}\label{eq:macc}
%\dot{M}_{\rm{acc}} = \frac{L_{\rm Bol}}{c^{2}\eta},
%\end{equation}
%
%\noindent where $\eta$ is the efficiency of accretion to convert rest mass energy into radiation and $c$ is the speed of light. We convert the absorption corrected 14--195 keV X-ray luminosity of $L_{X} = 2.15\times10^{43}$ ergs s$^{-1}$  from \citet{Ricci:2017aa} into $L_{\rm Bol}$ using the relation from \citet{Winter:2012yq} finding $L_{\rm Bol} = 1.4\times10^{44}$ erg s$^{-1}$. Assuming $\eta\approx0.1$ for a standard geometrically thin, optically thick accretion disk, we find a current $\dot{M}_{\rm{acc}} \approx 0.025$ M$_{\sun}$ yr$^{-1}$. This indicates that at relatively large scales, the mass inflow rate to the ring more than accounts for the mass accretion onto the SMBH and suggests a link between the feeding of the circumnuclear disk and the feeding of the AGN.

\subsection{The Inner Warped Disk}
%Kinematically, the molecular gas shows a clear signature of disk rotation with PA similar to the nuclear stellar ring. The velocity dispersion in the ring is $\sim10$ \kms{} and increases significantly to 100 \kms{} towards the AGN. The absolute LOS velocities also increase to 300 \kms{} and seem to change PA from 12\degr within the ring to $\sim-30\degr$ such that near the AGN the highest velocities are along the one edge of the holes. 
Inside the ring, both the stars and gas dramatically change their distribution and kinematics. The bottom row of Figure~\ref{fig:dustmaps} shows a zoom in of the inner 6.5\arcsec for the F160W image and $V-H$ dust map. We see inside from the contours of the continuum image and elongation along a P.A.$\sim$ 85\degr shown as a white line. This was first observed by \citet{Shaw:1993aa} who suggested the presence of a stellar nuclear bar possibly supported by the $x_{2}$ orbits of the main large scale bar. \citet{Wilson:1993aa} however also suggested it could just be scattered nuclear light off dust from the AGN. 

The dust map shows heavy extinction along an arc extending from the NE to SW directly across the nucleus. This is also the location of two strong concentrations of cold molecular gas that connect two sides of the ring. These two clumps straddle the very centre such that there is a lack of CO (2--1) emission at the location of the AGN and correspond to the double peaks seen in lower resolution CO (1--0) maps \citet{Petitpas:2002aa}. Perpendicular to the two clumps and the stream are also two distinct holes of CO (2--1) emission. These holes are aligned in the direction of the NLR and ionized gas outflow, and each have a radius of roughly 165 pc and cover a surface area of 8.6$\times10^{4}$ pc$^{2}$.

Using a 2\arcsec aperture and Equation~\ref{eq:lco}, we measure a molecular gas mass of $\sim4\times10^7$ M$_{\sun}$, or $\sim25$\% of the total molecular gas mass detected. Thus, while a large fraction of the molecular gas is contained within the rotating ring, a substantial amount seems to be experiencing irregular motion around the AGN.

The CO (2--1) line profiles near the centre contain multiple kinematic components which contribute to the increase of the measured velocity dispersion in the single Gaussian fits and are evidence for irregular motion. These kinematic components could be related to either the nuclear stellar bar or the ionized gas outflow so we performed a second fit of only the central 4\arcsec x 4\arcsec section of the ALMA cube. To fit each spaxel, we used the method described in \citet{Fischer:2017aa} that employs the Bayesian MultiNest algorithm \citep{Feroz:2009aa, Buchner:2014aa} to fit emission line profiles and calculate the evidence for a particular fit. In this way, by comparing the evidence from a single component fit to a double component fit, we can determine whether the second component is statistically necessary. This can be extended up to any number of components, and in this work we allow for a maximum of three components based on visual inspection of the ALMA cube. 

\begin{figure}
	\includegraphics[width=\columnwidth]{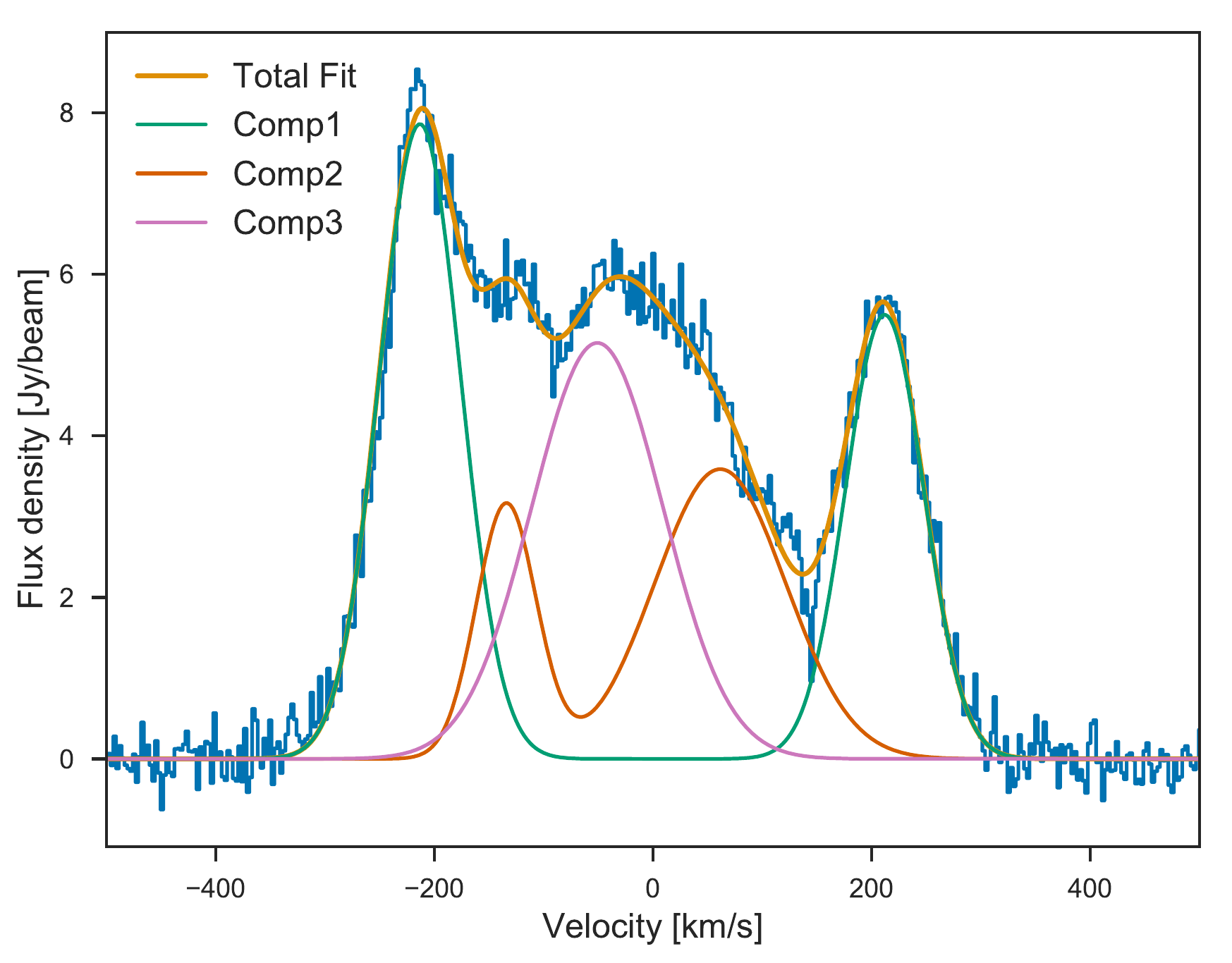}
	\caption{\label{fig:nuclear_co_spec} The integrated CO (2--1) spectrum within 2\arcsec of the nucleus (blue line) together with our fit using 5 kinematic components (orange line). We associate the two components with $\pm200$ \kms{} velocities (Comp1, green) and $\pm100$ \kms{} (Comp2, red) as single components. The final component near 0 \kms{} is Comp3 (pink).}
\end{figure}

\begin{figure*}
	\includegraphics[width=\textwidth]{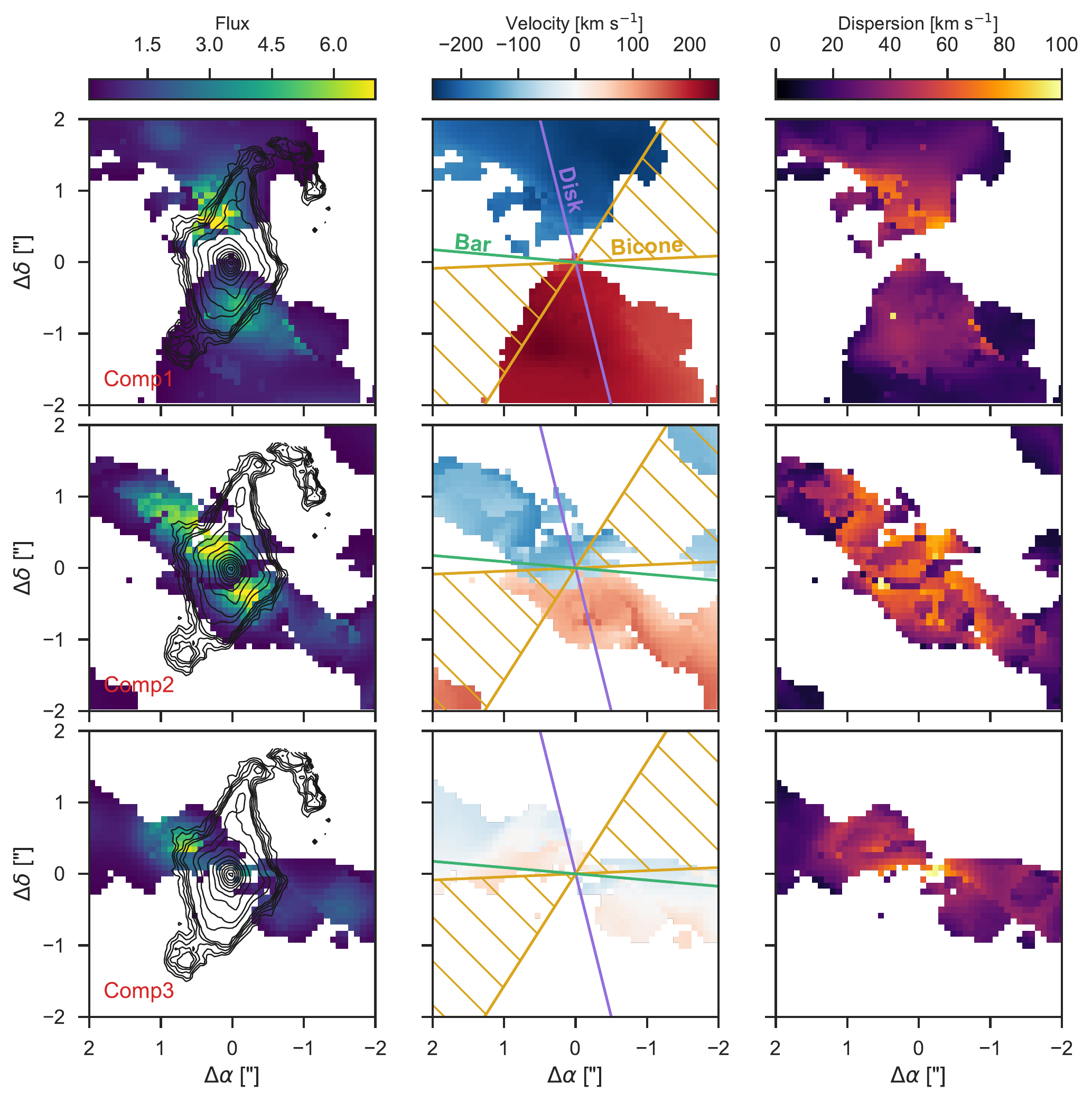}
	\caption{\label{fig:alma_multinest_fit}Results of our multi component fitting in the inner 4\arcsec of the ALMA cube. Each spaxel was fit with up to three velocity components. Each of the best fit components were then grouped into five groups according to their velocity. The first and fifth components are plotted together in the top panel as Comp1. The second and fourth components are plotted together in the middle panels as Comp2. The third component is plotted in the bottom panels. The black contours plot the flux distribution of \warmH{} from our SINFONI cube. In the middle panels, we also plot the kinematic major axis of the circumnuclear disk determined from our modelling of the larger scale CO (2--1) velocity field (purple), the ionized gas bicone region determined from the MUSE [OIII] flux map (gold), and the P.A. of the nuclear stellar bar shown also in Figure~\ref{fig:dustmaps} (green). }
\end{figure*}

A complication after fitting for multiple kinematic components is associating each velocity component within each spaxel with the ones in the other spaxels to form a coherent picture of the gas kinematics. To accomplish this, we first determined the likely independent components that existed in the region by fitting the integrated CO (2--1) spectrum within the central 2\arcsec. We found five distinct kinematic components that have velocities of roughly -200, -100, 0, 100, and 200 \kms{} which are shown together with the observed spectrum in Figure~\ref{fig:nuclear_co_spec}. Then, for each of the components for a single spaxel fit, we associated it with one of the five components based on which one it was closest to in velocity. In this way, we constructed flux, velocity, and dispersion maps for each of the five components. Finally, based on inspection of each of the maps it was clear that the first and fifth components as well as the second and fourth components are the opposite sides of the same physical component so we plot them together. Figure~\ref{fig:alma_multinest_fit} shows the results of our fits and reconstruction of each molecular gas component where Comp1 refers to the original first and fifth components ($\pm$200 \kms), Comp2 refers to the original second and fourth components ($\pm100$ \kms), and Comp3 is the original third component (0 \kms). 

Our analysis therefore suggests the presence of three kinematically distinct cold molecular gas components in the nuclear region of NGC 5728. The first, shown in the top row of Figure~\ref{fig:alma_multinest_fit}, is characterised by LOS velocities near 200 \kms{} and a P.A.$\sim$0\degr. The second, shown in the middle row, is characterised by LOS velocities around 100 \kms{} and a P.A.$\sim$45\degr, while the third component has LOS velocities near systemic and a kinematic P.A. that is unclear since the velocity gradient is weak but a flux P.A. near 90\degr.

We can also determine the mass of molecular gas that is contained within each kinematic component using again Equation~\ref{eq:lco} and the conversions given above. We find a total mass within 2\arcsec{} for Comp1, Comp2, and Comp3 of $1.6\times10^{7}$,  $1.6\times10^{7}$, and $7.7\times10^{6}$ M$_{\sun}$ respectively. Thus about 40\%, 40\%, and 20\% of the total nuclear molecular gas is contained in Comp1, Comp2, and Comp3. 

On all of the flux maps, we also plot the contours of \warmH emission from the SINFONI cube. On the velocity maps, we plot the kinematic major axis for the circumnuclear rotating disk based on our fit in Section~\ref{sec:disk_modeling}, the ionized gas bicone region based on the MUSE [OIII] map, and the major axis of the nuclear stellar bar based on Figure~\ref{fig:dustmaps}. 

Immediately it is clear that Comp1 likely corresponds to gas still in circular rotation in the inner part of the disk. However, it seems that as the molecular gas is rotating towards the bicone, the gas is being accelerated giving rise to the enhanced velocities seen here and in the residuals of the disk modelling (see Fig.~\ref{fig:disk_model}). However, as soon as the gas rotates into the bicone, all CO (2--1) emission disappears given none of the components we identified show any correspondence with the bicone. We discuss possible reasons for the disappearance of CO (2--1) emission in Section~\ref{sec:miss_mass}.

Comp2, however, seems to be aligned perpendicular to the bicone at least within the central 1\arcsec. Based on our later modelling of the ionized gas, we find an inclination for the bicone of 49\degr compared to an inclination for the disk of 43\degr. Since the bicone is directed almost straight into the disk, the molecular gas must warp by 90\degr in the inner nuclear region. Based on the P.A. shift for Comp2 compared to the larger scale disk and Comp1, we associate Comp2 with a warped inner rotating molecular disk that is contributing to shaping the bicone and likely providing the material necessary to fuel the AGN. Comp2 is further directly associated with the strong extinction seen in Figure~\ref{fig:dustmaps} and is therefore also providing the obscuration necessary to hide the BLR. It also has a relatively high velocity dispersion of $\sim80$ \kms{} also indicating a thick disk which are commonly found in the central few 100 pc of AGN \citep{Hicks:2009aa,Riffel:2011ab,Alonso-Herrero:2018aa}.

Comp3 perhaps is related to gas streaming along the nuclear bar. The velocities near 0 \kms{} therefore are due to the orientation of the bar in the plane of the sky. \citet{Petitpas:2002aa} were searching exactly for this molecular gas aligned with the nuclear stellar bar but were unable to detect it likely due to the low spatial resolution of their data. The thought was that nuclear gas in the bar would be an indication that the nuclear stellar bar would be driving gas further inward towards the AGN. However, we see here that while it does seem that there is molecular gas associated with the nuclear bar, the primary component that is driving gas to the centre is instead Comp2 which itself is arranged in a bar-like structure. This nuclear molecular gas bar then is significantly leading the stellar bar and in fact could be related to the $x_{2}$ orbits of the nuclear stellar bar which are perpendicular to the major axis of a bar. Hydrodynamical simulations do indicate that gaseous bars should lead stellar bars when both exist \citep{Friedli:1993aa,Shaw:1993aa} however note these simulations primarily consist of a single stellar and gaseous bar whereas for NGC 5728 we are observing a secondary nuclear and gaseous bar.  The three molecular gas components make for a seemingly chaotic environment around the AGN and is likely not a dynamically stable setup without collisions between gas clouds. It's possible though that this contributes to the variability of AGN where short-lived, unstable molecular gas environments quickly feed the AGN before breaking itself apart with the help of AGN feedback. 

The morphology of \warmH{} emission seems to be a combination of all three components. The distinct \textit{S}-shaped structure looks to follow the edges of Comp1 while the slight extension in the middle from NE to SW is either related to Comp2 or Comp3. The biggest difference between \warmH{} and CO (2--1) is that the peak of \warmH is directly located on the position of the AGN while the peaks of CO (2--1) straddle the centre. This combined with extended morphology following the edges of CO (2--1) emission suggests that the \warmH{} emission is produced by either X-ray heating from the AGN or shocks as the AGN driven outflow expands into the disk and is consistent with the conclusions of a study of NIR line emission in 62 AGN from long slit spectra \citep{Rodriguez-Ardila:2004aa,Rodriguez-Ardila:2005aa,Riffel:2013ck}. Indeed \citet{Durre:2018ab} show that this is the case through the use of H$_{2}$ excitation diagrams. 

\subsection{AGN Driven Outflow}\label{sec:outflow}

\subsubsection{NLR Morphology}
From the MUSE [OIII] maps (Figure~\ref{fig:zoom_maps}) we find the NLR in a biconical shape that seems to be ``piercing'' through the centre of the ring and has an apex exactly at the the location of the AGN. Particularly noticeable is the strong dip in [OIII] emission towards the NW that indicates obscuration by the nuclear star-forming disk. Combined with the fact that the SE emission is systematically brighter than the NW emission, the NLR is inclined such that the bicone is in front of the nuclear disk towards the SE and behind the disk towards the NW until it breaks through at larger radii. This interpretation is consistent with previous studies of NGC 5728 using HST narrow band imaging \citep{Wilson:1993aa} and lower spatial resolution IFU observations \citep{Davies:2016aa}.

%Using the location of the brightest star-forming clumps in the nuclear disk, we can fit an ellipse with a PA of -20\degr, a semimajor axis of 840 pc and an axis ratio of 0.67. Assuming the nuclear disk is perfectly circular, the axis ratio would indicate that the disk is inclined 48\degr with respect to the LOS. This corresponds well to the large scale, photometrically determined inclination of 55\degr.

We measure a biconical NLR extent of 1.7 kpc to the SE and 2.1 kpc to the NW. The central PA is roughly -60\degr and the full opening angle is $\sim$55\degr. The NLR emission, while extended out to kpc scales, is largely concentrated in two clumps on opposite side of the AGN. The brightest clump in the SE is only 130 pc away from the AGN while the NW one is about 230 pc. Separating the two clumps is another gap in emission that spatially corresponds to the Comp2 component of the cold molecular gas shown in Figure~\ref{fig:alma_multinest_fit} and suggests that it is the component that is obscuring the AGN and producing the biconical shape of the NLR. The reason that the gap in [OIII] emission does not fully align with the Comp2 molecular gas component is due to the inclination of the bicone. The SE side of the bicone is in front of Comp2 while the NW side is behind it such that the gap in [OIII] emission appears to the NW of the location of Comp2. What is abundantly clear though is that the cavities of cold molecular gas are filled with warm ionized gas.

\begin{figure}
	\includegraphics[width=\columnwidth]{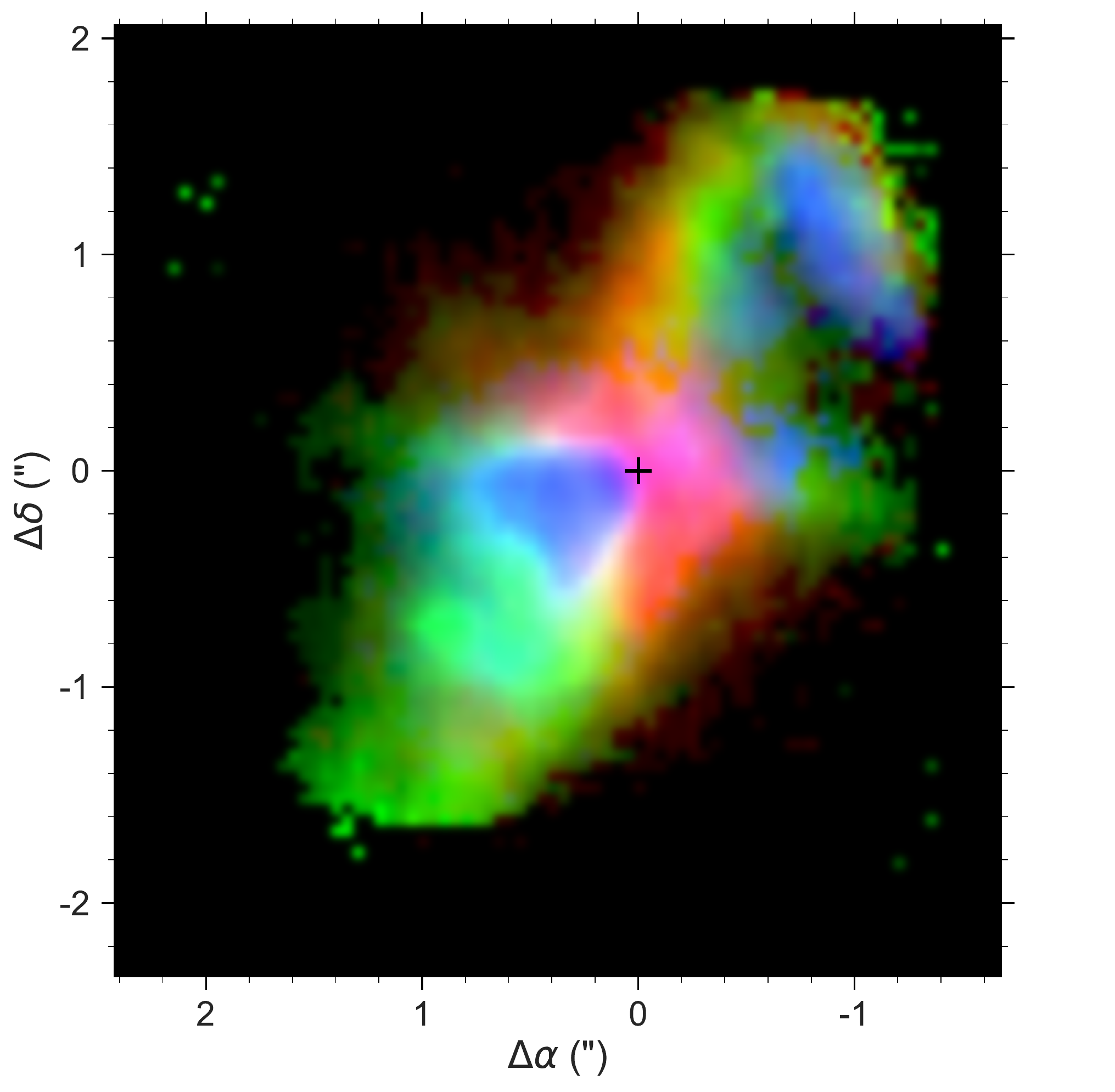}
	\caption{\label{fig:rgb_sinfoni} Three colour image of the central 4x4\arcsec of NGC 5728 from SINFONI. Red, green, and blue colours indicate emission from \warmH, [FeII], and [SiVI] respectively. The black cross marks the location of the AGN.}
\end{figure}

On smaller scales from the SINFONI data (Figure~\ref{fig:sinfoni_fits}), we find that the ionized emission is largely confined to the NLR showing again a biconical structure and extending from the SE to the NW, similar to the [OIII] emission from MUSE. The brightest emission for all three emission lines ([FeII], [SiVI], and \bry) is concentrated towards the SE, and with the higher spatial resolution of SINFONI, we locate it only 20 pc away from the AGN. Interestingly, [FeII] shows the same ``hook'' structure to the NW as the \warmH{} emission, except at locations inward from \warmH. Both [SiVI] and \bry{} instead are distributed in distinct clouds and have an overall shorter extent compared to [FeII] and \warmH. Figure~\ref{fig:rgb_sinfoni} highlights the spatial differences between the different phases of the gas with warm molecular gas (red) encompassing the partially ionized gas (green; [FeII]) which then encompasses the fully ionized gas (blue; [SiVI]). This observed ``stratification'' of the ionization structure of the gas around the nucleus strongly suggests that the AGN is the source of the ionization. This is also further evidence that the so-called ``Coronal Line Region'' is simply the inner part of the NLR where the ionizing radiation field is stronger and can produce lines with higher ionization potential. 

\subsubsection{Ionized Gas Kinematics}
%Kinematically, as morphologically, the ionized gas shows the presence of two components. On larger scales, the ionized gas velocity field seems to be rotating with the same kinematic P.A. as that of the stars and cold molecular gas as well as a fairly constant gas velocity dispersion similar to the molecular gas. This corresponds to the star-forming nuclear disk/ring encircling the AGN. Besides the nucleus, the velocity field does deviate from pure rotation towards the NW and SE where it looks as if there are two spiral arms. We interpret these deviations as the entry points of the gas to the nuclear disk/ring from the even larger scale bar. Gas streams radially along the bar from the NE and SW and fuels the nuclear disk/ring at these two points. The NE arm further spatially corresponds with the gap in the [OIII] emission seen in the flux map suggesting this gas is foreground to the NLR and not part of it. 

In the nucleus, and spatially corresponding to the brightest clumps of ionized gas emission, we find elevated velocities reaching +/- 400 \kms. The kinematic P.A. is along the same direction as the NLR and we see larger dispersions, although some of them are due to beam smearing and the presence of multiple kinematic components as in the cold molecular gas. All of this evidence points towards the presence of an AGN driven outflow, and the high spatial resolution SINFONI observations only further increase support for this interpretation. All of the SINFONI detected ionized emission lines show high velocities along the major axis of the NLR and dispersions reaching over 300 \kms (see Figure~\ref{fig:sinfoni_fits}). In particular, the [SiVI] line displays the largest velocities reaching 500 \kms. Both \bry{} and [FeII] show in general the same velocity structure albeit with slightly lower absolute velocities. This is in contrast to the \warmH kinematics which show much lower velocities and dispersions and suggests there is a lack of molecular gas in the outflow due to the AGN inflating a bubble in the disk such as the one seen in NGC 1068 \citep{May:2017aa}.

\subsubsection{Outflow Modeling}\label{sec:outflow_model}
Building on the outflow interpretation of the ionized gas kinematics, we used \textsc{DYSMALPY} to model and fit for the outflow parameters. As we did for modelling the rotating disk, we used the 2D velocity map as our input and specifically decided on the [SiVI] velocity map since it shows the strongest outflow signatures. Our biconical outflow model is similar to those that have been extensively used in the literature for modelling the kinematics of gas around nearby AGN based on both long slit data \citep[e.g.][]{Crenshaw:2000aa,Crenshaw:2000ab,Crenshaw:2010aa,Das:2005aa,Das:2006aa,Fischer:2013aa,Fischer:2014aa} as well as IFU data \citep[e.g.][]{Muller-Sanchez:2011aa, Bae:2017aa}. The model consists of two axisymmetric cones that share an apex at the location of the AGN. The cones have two opening angles, an inner ($\theta_{\rm inner}$) and outer ($\theta_{\rm outer}$) opening angle that define the walls of the cone and where the emission originates. The bicone also has an inclination ($i_{\rm bicone}$ where 0\degr{} indicates the cones are oriented along the line of sight) and position angle (P.A.$_{\rm bicone}$) on the sky with respect to the line of sight. 

As in \citet{Bae:2016aa}, we assume the flux is exponentially decreasing with radius according to $F(r) = Ae^{-\tau r/r_{end}}$ where $r_{\rm end}$ is the radial edge of the bicone and $\tau$ controls the speed at which the flux falls off. Given we are only fitting the velocity map, we fixed $\tau$ to be 5 which is the value found for many nearby AGN. Finally, the model allows for different velocity profiles that change as a function of radius. We chose a linearly increasing profile that reaches a maximum ($v_{\rm max}$) at a specific radius ($r_{\rm turn}$) then decreases linearly until $r_{\rm end}$. This assumes a radially symmetric profile such that $r_{\rm end} = 2r_{\rm turn}$. As with modelling the rotating disk, we determined the best fit parameters using MCMC with 800 walkers, 100 burn steps, and 100 sampling steps.

\begin{figure*}
	\includegraphics[width=\textwidth]{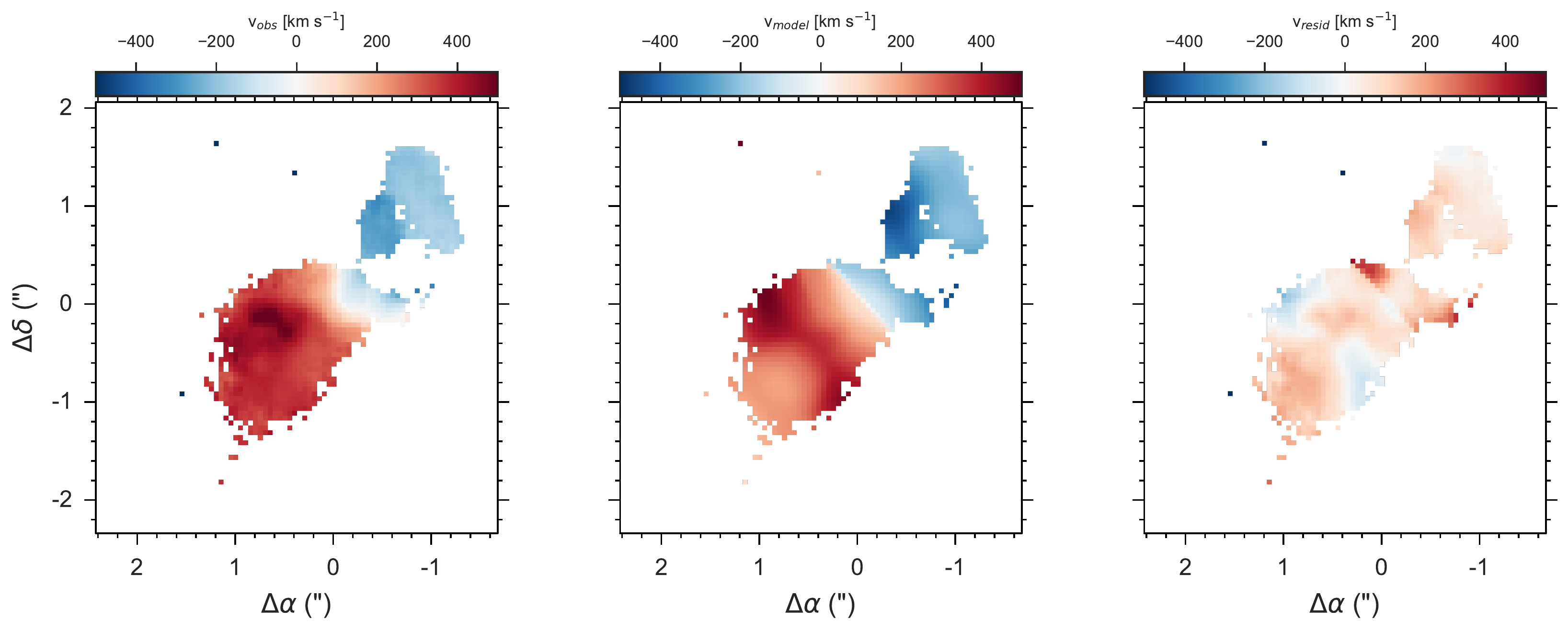}
	\caption{\label{fig:outflow_model}Results of fitting the [SiVI] velocity field with a biconical outflow model. The observed velocity field is shown in the left panel, the best fit model in the middle panel, and the residuals in the right panel. }
\end{figure*}

Figure~\ref{fig:outflow_model} shows the results of our outflow fitting. We find best fit outflow parameters of $\theta_{\rm inner} = 18\pm1\degr$, $\theta_{\rm outer} = 23\pm1\degr$$i_{\rm bicone} = 49\pm2\degr$, P.A.$_{\rm bicone} = 43\pm2\degr$, $r_{\rm end} = 540\pm20$ pc, and $v_{\rm max} = 737\pm25$ \kms based on the median and standard deviations of the resulting posterior distributions. %The model is not a perfect representation of the velocity structure of the ionized gas with residuals reaching over 100 \kms{} in some pixels. However, the best fit bicone model does reproduce the overall general radial velocity profile. 

%\begin{figure*}
%	\includegraphics[width=\textwidth]{figures/combined_model_with_halpha_oiii}
%	\caption{\label{fig:combined_disk_outflow} Comparison between the combined disk and outflow model (left panel) from our fitting of the cold molecular gas and [SiVI] emission with the [OIII] (middle panel) and \halpha{} (right panel) velocity fields. The black ellipse outlines the location of the circumnuclear ring as seen in Figure~\ref{fig:circum_ring}.}
%\end{figure*}

%Figure~\ref{fig:combined_disk_outflow} then shows the combination of our best fit disk model from the cold molecular gas with the best fit outflow model from [SiVI] and compares it to the velocity fields of [OIII] and \halpha{} from MUSE. The model was appropriately adjusted to match the pixel scale and PSF of MUSE. To create the combined model, we manually adjusted the relative intensity scaling of the disk and outflow component until a satisfactory model velocity field was achieved. Again, the combined model is not a perfect representation of the larger scale ionized gas kinematics but overall it captures the high velocity central peaks and general rotation on larger scales. As mentioned previously, we are not able to model the kinematic ``arms'' which are likely due to gas inflowing into the nucleus from the bar.

Comparing the spatial orientations of the disk and outflow, we see that our initial interpretation that the NLR is in front of the disk towards the SE and behind the disk towards the NW is correct. The disk is inclined at an angle of 43\degr{} with respect to the LOS, while the outflow is inclined by 49\degr. However, an outer half opening angle of 23\degr{} places the front side of the NLR at only an angle of 26\degr{}, well in front of the disk. The similarity of the inclinations for both the disk and the outflow further shows that the NLR is completely intersecting the disk. The turnover radius for the outflow is only 270 pc which explains why we only observe enhanced velocities very near to the centre in the MUSE maps. Thus, while the NLR extends out to 2 kpc only the very central regions are outflowing and the rest is simply AGN excited gas within the disk. 

\subsection{Outflow Energetics}
In this section, we explore the ionized gas energetics within the outflow. Because of the short extent of the outflow, we primarily use the diagnostics available from the SINFONI cube to spatially resolve the outflow properties. In particular, we want to measure the mass outflow rate, mechanical energy rate, and momentum rate as a function of radius and compare to the energy and momentum injection from the central AGN. 

Similar to our analysis of the electron density radial profile in Appendix~\ref{sec:electron_density}, we binned the SINFONI cube using annular apertures along the SE redshifted half of the NLR given extinction is not as much of a factor. For the SINFONI data, we chose an annular width of 0.075\arcsec{} which is half the PSF size and extended out to 1.65\arcsec. Within each aperture, we fit the [SiVI] and \bry{} line profile with either 1 or 2 Gaussian components, choosing between them based on $\log$(evidence) as we did for the very central regions of the ALMA data. If the profile needed two components, we calculated a flux-weighted average velocity since we assume that all of the emission in the inner regions is related to the outflow.

\begin{figure*}
	\includegraphics[width=0.9\textwidth]{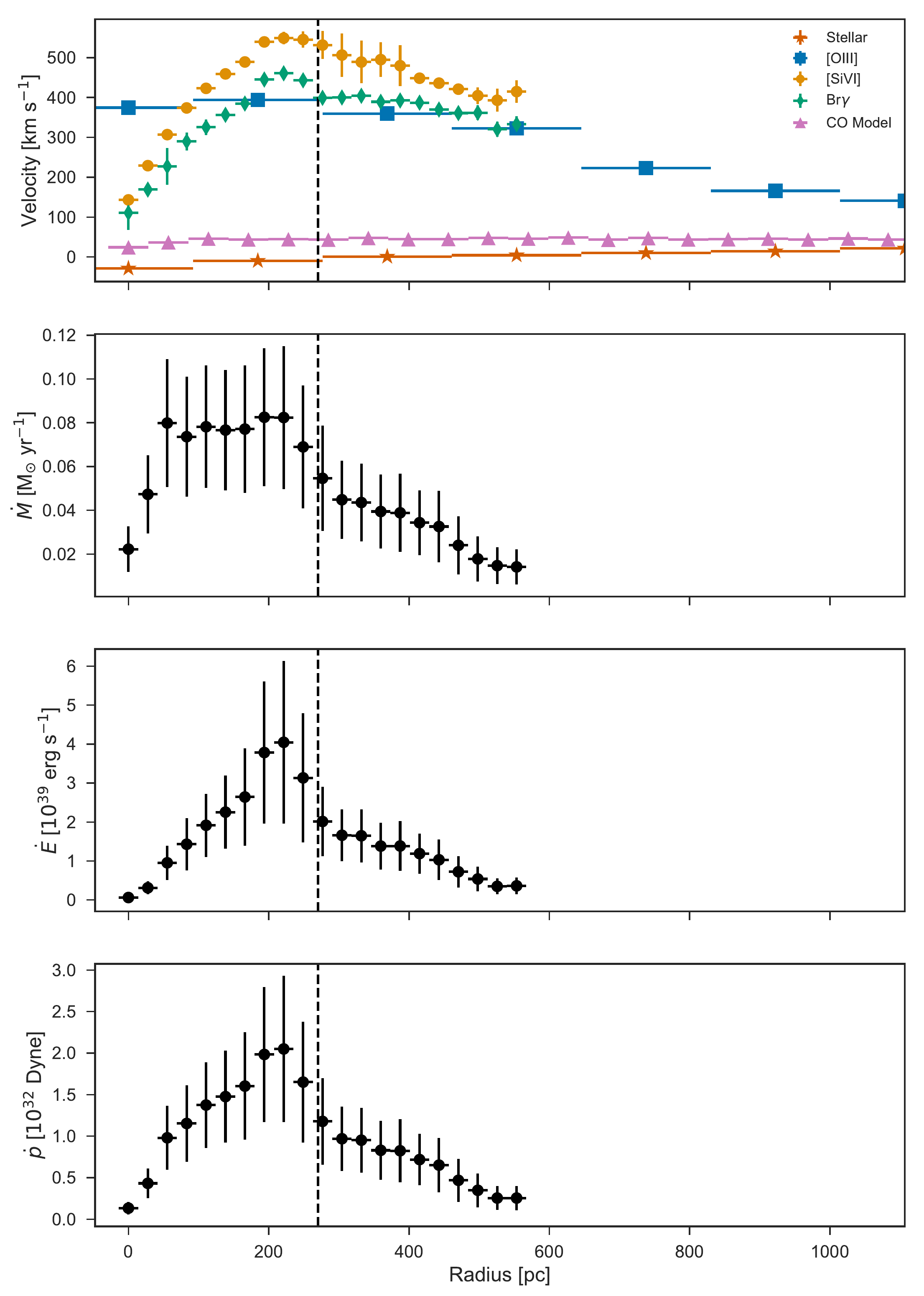}
	\caption{\label{fig:outflow_energetics}Radial profiles of the outflow velocity (top panel), mass outflow rate (second panel), outflow kinetic luminosity (third panel), and momentum rate (bottom panel) determined from concentric annuli spanning the redshifted side of the NLR. Velocity profiles are shown for [SiVI], \bry, [OIII], the stars, and CO (2--1). The CO (2--1) velocities were inferred from the best fit rotating disk model due to the lack of CO (2--1) emission in the bicone. The dashed vertical line indicates the best fit turnover radius from our 2D outflow modelling of the [SiVI] velocity map.}
\end{figure*}

The upper panel of Figure~\ref{fig:outflow_energetics} plots the resulting velocity profile for both [SiVI] and \bry{} with all radii and velocities deprojected based on the inclination from the best fit 2D model from Section~\ref{sec:outflow_model}. Both ionized gas lines show essentially the same profile, implying that both lines are tracing the same physical components of gas. The narrow and broad components are also both increasing with radius up to $\sim$250 pc, then decreasing toward larger radii just as we modelled previously. The general consistency of both the narrow and broad component suggests that these are also tracing the same gas and merely reflect the underlying velocity distribution of the gas. We also show the few inner bins of the [OIII] velocity profile from MUSE which traces roughly the same shape except a bit more flattened due to the larger beam size of the MUSE observation. We further show the velocity profiles from the stars and the best fit rotating disk model used to fit the CO (2--1) velocity map. These both show a nearly flat profile around 0 \kms{} since the bicone axis is aligned nearly along the minor axis of the disk.

Because \bry{} and [SiVI] originate in the same gas we can measure the ionized gas mass using the total flux of the \bry{} line assuming Case B recombination gas conditions. Further since \bry{} occurs in the NIR, dust extinction is negligible. We convert from \bry{} flux to gas mass within each aperture using the formula given in \citet{Storchi-Bergmann:2009aa}:

\begin{equation}\label{eq:bry_to_mass}
M_{\scriptscriptstyle \rm HII} = 3\times10^{19}\Biggl(\frac{F_{\scriptscriptstyle \rm Br \gamma}}{\rm erg\,cm^{-2}\,s^{-1}}\Biggr)\Biggl(\frac{D}{\rm Mpc}\Biggr)^{2}\Biggl(\frac{n_{e}}{\rm cm^{-3}}\Biggr)^{-1}
\end{equation} 

\noindent where $D$ is the luminosity distance in Mpc and $n_{e}$ is the electron density. $M_{\scriptscriptstyle \rm HII}$ then is the total ionized gas mass in solar units. For $n_{e}$, we assume a constant electron density within the SINFONI FOV and use the value obtained from our analysis of the auroral and trans-auroral lines (1550 cm$^{-3}$; see Appendix~\ref{sec:electron_density}). We can then calculate the mass outflow rate, mechanical luminosity, and outflow momentum rate using the following standard equations:

\begin{equation}\label{eq:mass_outflow}
\dot{M}_{\rm{out}} = \left(\frac{M_{\scriptscriptstyle \rm HII}v_{\rm{out}}}{\delta r}\right),
\end{equation}

\begin{equation}\label{eq:energy_rate}
\dot{E}_{\rm{out}} = \frac{1}{2}\dot{M}_{\rm{out}}v_{\rm{out}}^{2},
\end{equation}

\begin{equation}\label{eq:momentum_rate}
\dot{p}_{\rm{out}} = \dot{M}_{\rm{out}}v_{\rm{out}},
\end{equation}

\noindent where $\delta r$ is the width of the annular apertures. The bottom three panels of Figure~\ref{fig:outflow_energetics} show our calculations of $\dot{M}_{\rm{out}}$, $\dot{E}_{\rm{out}}$, and $\dot{p}_{\rm{out}}$ as a function of distance from the AGN. $\dot{M}_{\rm{out}}$, $\dot{E}_{\rm{out}}$, and $\dot{p}_{\rm{out}}$ all show peak values around 200 pc and peak values of 0.08 M$_{\sun}$ yr$^{-1}$, $4\times10^{39}$ erg s$^{-1}$, and $2\times10^{32}$ Dyne. We note that our outflow measurements disagree from those presented in \citet{Durre:2018aa} who reported $\dot{M}_{\rm{out}} = 38$ M$_{\sun}$ yr$^{-1}$, almost 500 times greater than ours. We attribute this difference to two factors: 1) they assumed an electron density of 100 cm$^{-3}$ compared to our measured value of 1000 cm$^{-3}$ and 2) they summed up the outflow rate determined within each single spaxel over the entire SINFONI FOV. For mass outflow rates, this is not correct as the rate needs to be calculated within a common radius.

The peak mechanical energy in the outflow is well below the AGN luminosity, $L_{\rm Bol} = 1.4\times10^{44}$ erg s$^{-1}$, only reaching 0.003\% of $L_{\rm Bol}$. Similarly, the ratio of the momentum rate to the radiation momentum rate from the AGN ($L_{\rm Bol}$/$c$) is only 0.043. These properties all suggest that the outflow currently is not energy driven but rather momentum driven given their extremely low values \citep[e.g.][]{Faucher-Giguere:2012aa,Zubovas:2012aa}. 

While the values seem quite low, we can compare them to the properties of outflows found across literature. \citet{Fiore:2017aa} recently compiled a comprehensive set of outflow energetics from various studies over the years spanning a large range of AGN luminosity. They used this data to establish clear AGN wind scaling relations, in particular between the mass outflow rate and outflow power with the AGN bolometric luminosity in all phases of the gas in the outflow. 

\begin{figure}
	\includegraphics[width=\columnwidth]{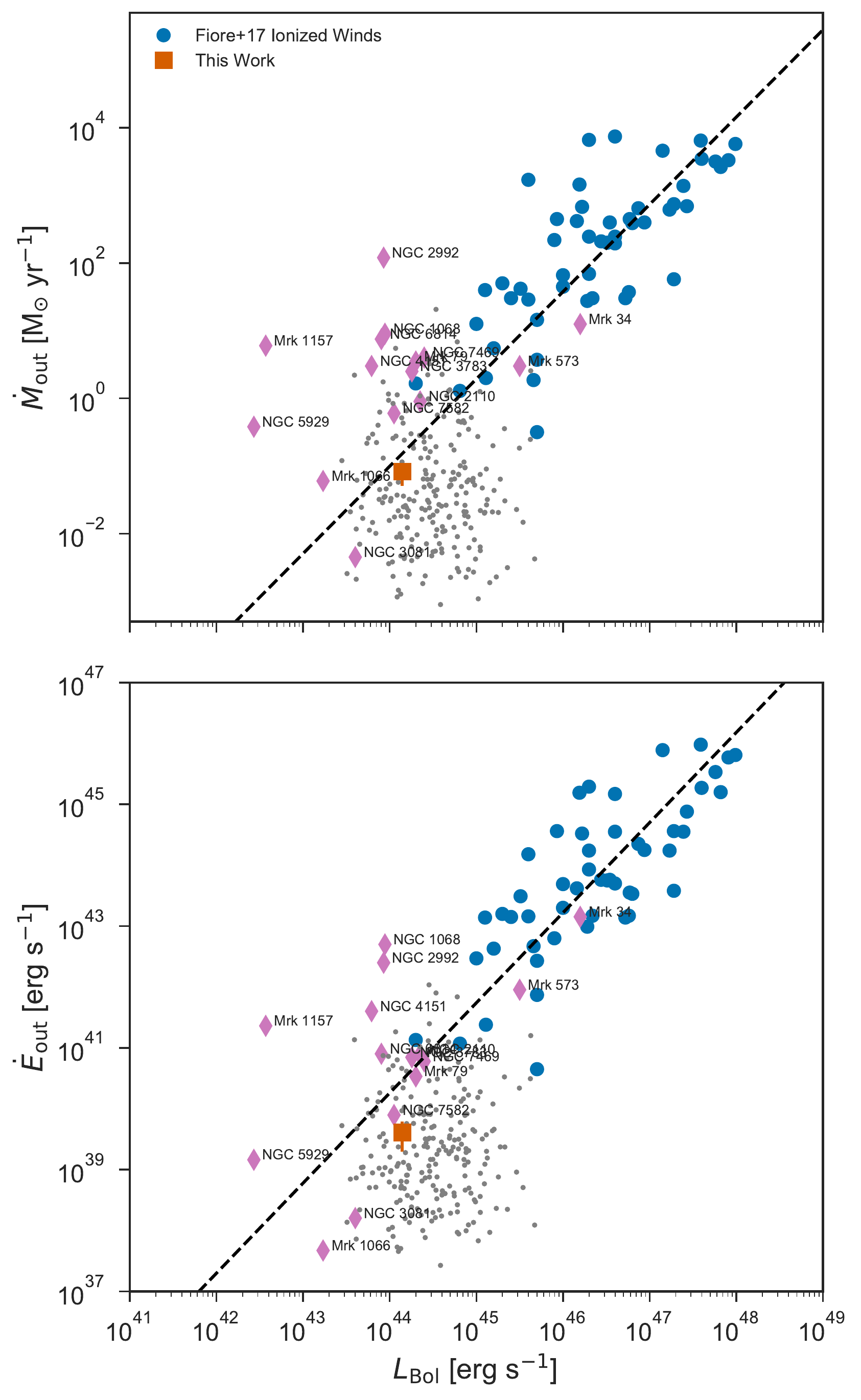}
	\caption{\label{fig:comp_fiore}Correlations between the mass outflow rate ($\dot{M}_{\rm{out}}$; top panel) and kinetic outflow power ($\dot{E}_{\rm{out}}$, bottom panel). Blue points are the data compiled by \citet{Fiore:2017aa} and the orange square plots our values for NGC 5728. Purple diamonds are compiled from various literature sources for more modest AGN luminosities. See the text for the specific references for each of the AGN. Gray points are the type 2 AGN studied in \citet{Baron:2019aa}. The black dashed lines show the best fit correlations from \citet{Fiore:2017aa}.}
\end{figure}

In Figure~\ref{fig:comp_fiore}, we compare our values of $\dot{M}_{\rm{out}}$ and $\dot{E}_{\rm{out}}$ for NGC 5728 with the ionized wind data from \citet{Fiore:2017aa}. NGC 5728 lies very close to the correlations and the low values of $\dot{M}_{\rm{out}}$ and $\dot{E}_{\rm{out}}$ seem to be simply due to the relatively modest AGN luminosity. With only one point, we cannot definitively determine if the correlations presented in \citet{Fiore:2017aa} actually extend to lower luminosities. Therefore, we compiled more values for $\dot{M}_{\rm{out}}$ and $\dot{E}_{\rm{out}}$ with a focus on AGN with $L_{\rm Bol} \la 10^{46}$ erg s$^{-1}$. We found data for 15 AGN: NGC 1068, NGC 2992, NGC 3783, NGC 6814, NGC 7469 \citep{Muller-Sanchez:2011aa}, NGC 4151 \citep{Crenshaw:2015aa}, Mrk 573 \citep{Revalski:2018aa}, Mrk 34 \citep{Revalski:2018ab}, NGC 7582 \citep{Riffel:2009aa}, Mrk 1066 \citep{Riffel:2011ab}, Mrk 1157 \citep{Riffel:2011aa}, Mrk 79 \citep{Riffel:2013ab}, NGC 5929 \citep{Riffel:2015aa}, NGC 2110 \citep{Schnorr-Muller:2014aa}, and NGC 3081 \citep{Schnorr-Muller:2016aa}. Finally, we also include the sample of 234 type 2 AGN studied in BN19 using single integrated SDSS spectra and dust SED fitting. 

%While there is substantial scatter towards lower $L_{\rm Bol}$, it does seem as if the \citet{Fiore:2017aa} relations hold indicating that the radiation strength of the AGN is the main determinant of the strength of the outflow and thus its effect on the host galaxy. We note that for the AGN compiled from other works, we made no attempt to homogenise the methodology used to define these quantities as \citet{Fiore:2017aa} did. Further each of the works uses different values for the electron density except for NGC 4151, Mrk 573, and Mrk 34 where detailed photoionization modelling was performed to measure the mass in the outflow. Normalising the analysis methods and ensuring accurate estimates of the electron density would likely decrease the large scatter seen in Figure~\ref{fig:comp_fiore}. 

The inclusion of both single source literature compilation and the BN19 sample shows that while NGC 5728 lies close to the \citet{Fiore:2017aa} correlations, there is likely substantial scatter at lower luminosities. That said, combining all of these samples together can be dangerous given the widely varying methods and data quality used to measure these properties. In particular, while \citet{Fiore:2017aa} renormalized all outflow properties to a constant electron density, BN19 showed that object by object determined densities can change the mass outflow rate considerably and thus weaken the mass outflow rate correlation. %Finally, it is unclear whether these correlations display anything more than the well known NLR luminosity vs. AGN luminosity correlation given that the method to calculate mass outflow rate is essential $L_{\rm line}v_{\rm out}/r_{out}$. 
What is needed to confirm these correlations is a large, spatially resolved sample of AGN where all of these properties can be properly and systematically measured.

%We can further compare the ionized gas outflow with the potential molecular gas outflow seen in the residuals of our disk modelling of CO (2--1) emission. Recall in the inner regions of the disk we observed residuals consistent with radial outflow indicating a peak outflow velocity of 40 \kms{} and molecular mass outflow rate of 5 M$_{\sun}$ yr$^{-1}$. Therefore while the outflow velocity of the molecular gas is much lower than the ionized gas, the mass outflow rate is 60 times higher, consistent with what is commonly seen other moderate luminosity AGN \citep[e.g.][]{Fiore:2017aa}.

\subsection{AGN Feedback in NGC 5728}
With such a low mass outflow rate and outflow power, it is unlikely that the AGN driven wind is effectively suppressing star formation in NGC 5728 and affecting its evolution. With a total molecular gas mass in the circumnuclear disk of 10$^{8.1}$ M$_{\sun}$, it would take 217 Myr to completely clear the disk of its gas even including the potential molecular gas outflow. Using \textit{Herschel} photometry and SED fitting, \cite{Shimizu:2017aa} estimated a global SFR of 2.1 M$_{\sun}$ yr$^{-1}$. Combining this with $\dot{M}_{\rm{out}}$, the total gas depletion time still 47 Myr. However, this is under the assumption that all of the outflowing gas can escape the gravitational potential of the galaxy. Following \citet{Rupke:2002aa}, we can estimate the escape velocity under the assumption of a truncated isothermal sphere, $v_{\rm esc}(r) = \sqrt(2)v_{\rm circ}[1 + \ln(r_{\rm max}/r)]^{1/2}$. We assume an $r_{\rm max} = 100$ kpc and use $v_{\rm circ} = 320$ \kms{} from \citet{Rubin:1980aa}. At 250 pc, the radius of peak outflow velocity, $v_{\rm esc} = 1200$ \kms, far above the outflow velocities of either the ionized or molecular gas. Outflows in moderate luminosity AGN do not seem to have the energy to escape their host galaxies as this has been observed in many other systems \citep[e.g.][]{Davies:2014aa,Fischer:2017aa,Herrera-Camus:2018aa}. 

%Further it does not seem as if the outflow reaches anywhere beyond the inner few 100 pc of the disk and in fact we see clear deceleration back to disk rotation at $\sim1$ kpc. %Therefore while there is some stirring up of the gas, it is likely only being lifted up a little bit above the disk before falling back down.

We can further compare the mass outflow rate with the mass flow rate through the bicone. Using the best fit parameters for the outflow, the total surface area covered by the bicone is 0.19 kpc$^{2}$. From our disk modelling we find an average rotational velocity of the molecular gas of 273 \kms{} through the bicone. Combined with the average surface mass density of $55$ M$_{\sun}$ pc$^{-2}$, we calculate that molecular gas flows through the bicone at a rate of $\sim$ 9.5 M$_{\sun}$ yr$^{-1}$, about 16x higher than the mass outflow rate. Essentially, the gas that is entering the bicone is not being ejected at rates large enough that would substantially disturb the structure and dynamics of the disk. This is especially true when we take into account the fact that the molecular gas is only moving radially at speeds of 40 \kms{} which means that by the time it exits the bicone it has only moved about 55 pc.

%Instead, the AGN must increase in power substantially to start driving significant amounts of gas out. Assuming the \citet{Fiore:2017aa} correlations hold, a jump to a bolometric luminosity of 10$^{48}$ erg s$^{-1}$ would increase the mass outflow rate to roughly 10$^{3.5}$ M$_{\sun}$ yr$^{-1}$. This would drastically decrease the depletion time to only 4$\times10^{4}$ yr. To reach this luminosity, the AGN would need a large increase in its accretion rate. To reach up to $L_{\rm Bol} = 10^{48}$ erg s$^{-1}$ would require an $\dot{M}_{\rm{acc}} = 177$ M$_{\sun}$ yr$^{-1}$. However, based on a SMBH mass of 10$^{7.36}$  from megamaser observations \citep{Braatz:2015aa}, the Eddington luminosity ($L_{\rm Edd}$ ) is only $\approx 10^{46}$ erg s$^{-1}$ which would reduce the maximum $\dot{M}_{\rm{out}}$ to 10 M$_{\sun}$ yr$^{-1}$ and increase the depletion time to about 13 Myr. 

Finally, we can compare the mass outflow rate to the current mass accretion rate onto the SMBH using the following equation:

\begin{equation}\label{eq:macc}
\dot{M}_{\rm{acc}} = \frac{L_{\rm Bol}}{c^{2}\eta},
\end{equation}

\noindent where $\eta$ is the efficiency of accretion to convert rest mass energy into radiation and $c$ is the speed of light. %We convert the absorption corrected 14--195 keV X-ray luminosity of $L_{X} = 2.15\times10^{43}$ ergs s$^{-1}$  from \citet{Ricci:2017aa} into $L_{\rm Bol}$ using the relation from \citet{Winter:2012yq} finding $L_{\rm Bol} = 1.4\times10^{44}$ erg s$^{-1}$. 
Assuming $\eta\approx0.1$ for a standard geometrically thin, optically thick accretion disk, we find a current $\dot{M}_{\rm{acc}} \approx 0.025$ M$_{\sun}$ yr$^{-1}$. Therefore, even the ionized gas mass outflow rate is three times the current mass accretion rate onto the SMBH implying that the majority of gas near the AGN does not end up fuelling it. 

\subsection{Missing molecular gas?}\label{sec:miss_mass}
The most striking evidence of AGN feedback is the presence of what appears to be large ``holes'' of cold molecular gas as probed by CO (2--1) emission that are co-located exactly along the outflow and NLR. Therefore, it's reasonable to interpret that the AGN driven outflow has completely evacuated these regions of the disk of molecular gas. However, these holes do not exist in maps of the ionized gas and in fact it is these regions where emission from ionized gas is strongest. If an AGN driven outflow had indeed evacuated these regions, we should expect it to drive out all phases of gas.

A plausible explanation for these holes is instead that CO (2--1) is not accurately tracing the presence molecular gas in the NLR. Under standard ISM conditions, CO (2--1) would be a reliable tracer of the molecular gas mass, however in the presence of the strong radiation field of an AGN, it could be that the CO (2--1) line in particular or all CO lines are suppressed. Whereas in a normal star-forming disk, molecular gas is primarily heated by nearby OB stars with FUV photons dominating the process, near an AGN, X-rays will dominate over FUV photons and control the heating and chemical composition of the gas producing a so-called X-ray dissociation region (XDR). 

This is being increasingly observed in local AGN. Both \citet{Rosario:2019aa} and \citet{Feruglio:2019aa} reported recently on a lack of CO (2--1) emission near the central AGN that is instead filled with \warmH{} and ionized gas emission. These cavities of CO emission are further cospatial with hard X-ray emission, confirming that X-ray radiation is suppressing the CO emission. While we observe the same effect at the very centre of the galaxy corresponding to the location of the AGN, we do not observe \warmH{} emission filling the larger scale CO (2--1) holes that correspond to the bicone. Instead, \warmH{} is only lining the edges of the bicone where gas is entering the bicone. It's possible then that the inner NLR is devoid of molecular gas and a continuation of the stratification we observe in the SINFONI line emission continues on larger scales, i.e. that cold molecular gas only exists outside the bicone region with hotter and hotter gas becoming more prevalent towards the inner NLR. To confirm the lack of molecular gas in the NLR, however, observations of higher-J CO lines are needed.

\section{Summary \& Conclusions}
We have presented a comprehensive analysis of multiwavelength datasets for the nearby Seyfert 2 galaxy, NGC 5728. Using primarily HST imaging and MUSE, SINFONI, and ALMA cubes, we measured and analysed the distribution and kinematics of the stars, ionized gas, and molecular gas. Our results and conclusions are summarised as follows:

\begin{itemize}
    
    \item Prominent dust lanes are observed extending along the major axis of the large stellar bar from the outer disk to the circumnuclear ring. These are likely coincident with the primary shocks produced by the axisymmetric bar potential and driving inflow to the circumnuclear region.
    
    \item The circumnuclear ring is observed as a ring of low stellar velocity dispersion, bright \halpha{} clumps, and CO (2--1) emission indicating the presence of on ongoing star formation and young stars. This star formation is likely induced by the build up of gas at the Inner Lindblad Resonance of the primary bar. The kinematics of the ring are well fit by circular rotation, however we also seen signs of inflow into the ring where the ring and dust lanes meet at a rate of 1 M$_{\sun}$ yr$^{-1}$. %This is more than enough to support the current 0.025 M$_{\sun}$ yr$^{-1}$ accretion rate of the AGN. 
    
    \item Inside the ring, we find a three distinct kinematic components of the molecular gas corresponding to gas rotating with the ring, gas inflowing to the AGN, and gas following the nuclear stellar bar. The gas inflowing to the AGN is distributed across the nucleus at the location of heavy extinction and is the source of obscuration for the AGN due to its perpendicular orientation compared to the NLR. 
    
    \item The AGN is driving a weak outflow primarily seen in ionized gas with a mass outflow rate of 0.08 M$_{\sun}$ yr$^{-1}$ that only reaches to radii of 250 pc before decelerating down to stellar rotation velocities. There are hints of molecular outflow in residuals of the disk modelling indicating a molecular mass outflow rate of 1 M$_{\sun}$ yr$^{-1}$ but only 40 \kms velocity. We determine that the outflow is unlikely to be largely disturbing the structure of the circumnuclear disk.
    
    \item We observe cavities of CO emission co-spatial with the AGN outflow that could either be explained by a deficiency of molecular gas in the NLR or a suppression of CO emission by the hard X-ray radiation of the AGN.
    
\end{itemize}

Overall, for NGC 5728, we find signatures of both feeding and feedback of the AGN. However, feeding of both the AGN and central region of the galaxy is currently the dominant process. NGC 5728 would need a large increase to its accretion rate to boost the energetics of its outflow and severely disrupt its evolution.

\section*{Acknowledgements}
We thank the anonymous referee for helpful comments and suggestions which improved the paper. We thank Cheng-Yu Kuo from the Megamaser Cosmology Project for sharing the coordinates of the AGN in NGC 5728. TTS thanks Travis Fischer for sharing and discussing his line decomposition software package as well as discussing the results of this paper. This paper makes use of the following ALMA data: ADS/JAO.ALMA\#2015.1.00086.S. ALMA is a partnership of ESO (representing its member states), NSF (USA) and NINS (Japan), together with NRC (Canada) and NSC and ASIAA (Taiwan) and KASI (Republic of Korea), in cooperation with the Republic of Chile. The Joint ALMA Observatory is operated by ESO, AUI/NRAO and NAOJ. This research has made use of the NASA/IPAC Extragalactic Database (NED), which is operated by the Jet Propulsion Laboratory, California Institute of Technology, under contract with the National Aeronautics and Space Administration. This research has made use of NASA's Astrophysics Data System Bibliographic Services. This research made use of Astropy,\footnote{http://www.astropy.org} a community-developed core Python package for Astronomy \citep{Astropy:2013ek, Astropy:2018dk}.
%%%%%%%%%%%%%%%%%%%%%%%%%%%%%%%%%%%%%%%%%%%%%%%%%%



\bibliographystyle{mnras}
\bibliography{ngc5728_bib} % if your bibtex file is called example.bib

%%%%%%%%%%%%%%%%%%%%%%%%%%%%%%%%%%%%%%%%%%%%%%%%%%

%%%%%%%%%%%%%%%%% APPENDICES %%%%%%%%%%%%%%%%%%%%%

\appendix

\section{Full FOV Emission Line Fits}

\begin{figure*}
    \includegraphics[width=\textwidth]{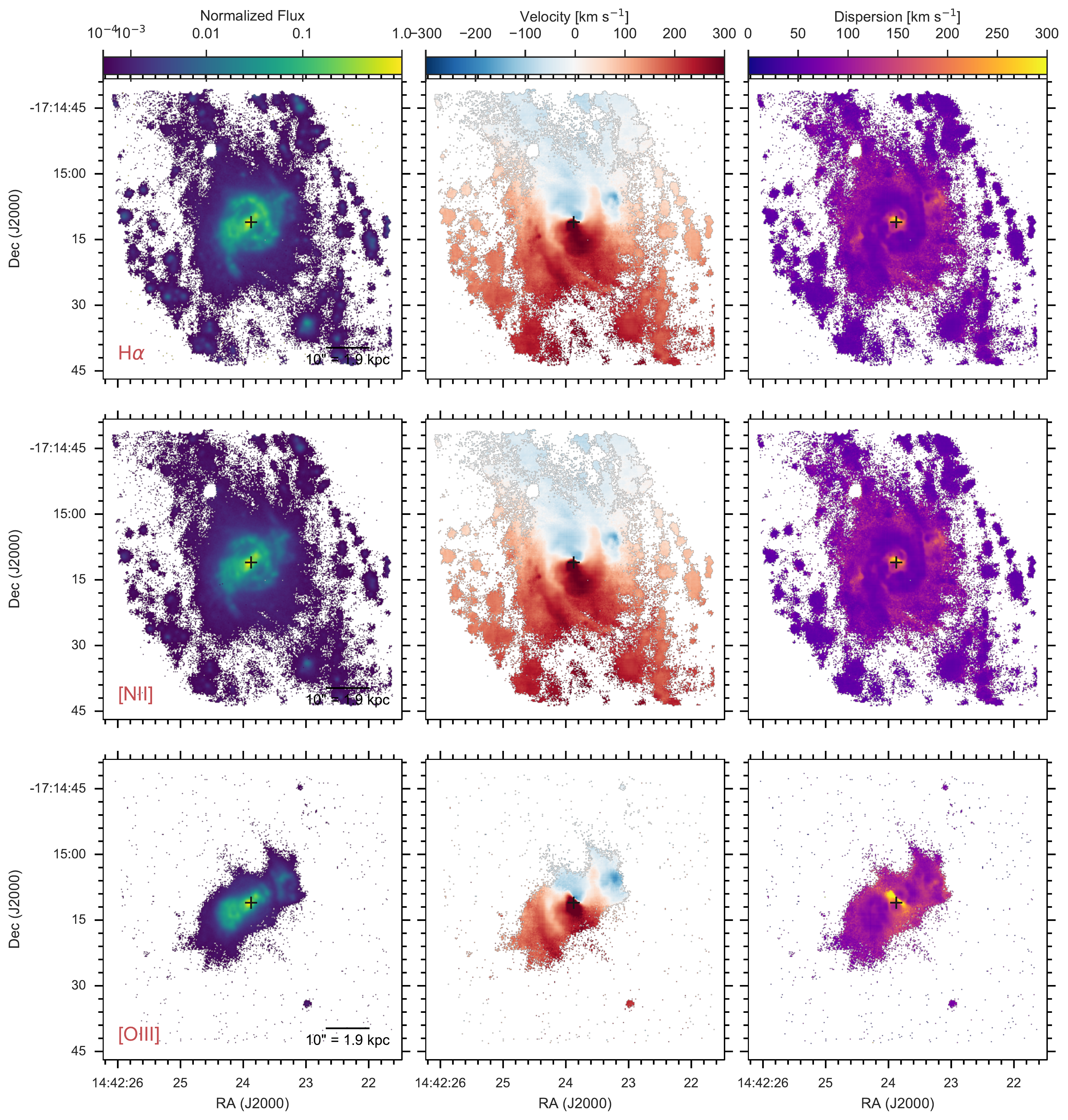}
    \caption{Single Gaussian fits to the \halpha, [NII], and [OIII] lines detected in the MUSE cube. Left column shows the integrated flux normalized to the maximum flux, middle column shows the velocity, and the right column shows the velocity dispersion. Velocity dispersions have been corrected for instrumental line broadening. In all frames, North is up and East is to the left. The black cross locates the position of the AGN.\label{fig:muse_1_fits}}
 \end{figure*}
 
 \begin{figure*}
    \includegraphics[width=\textwidth]{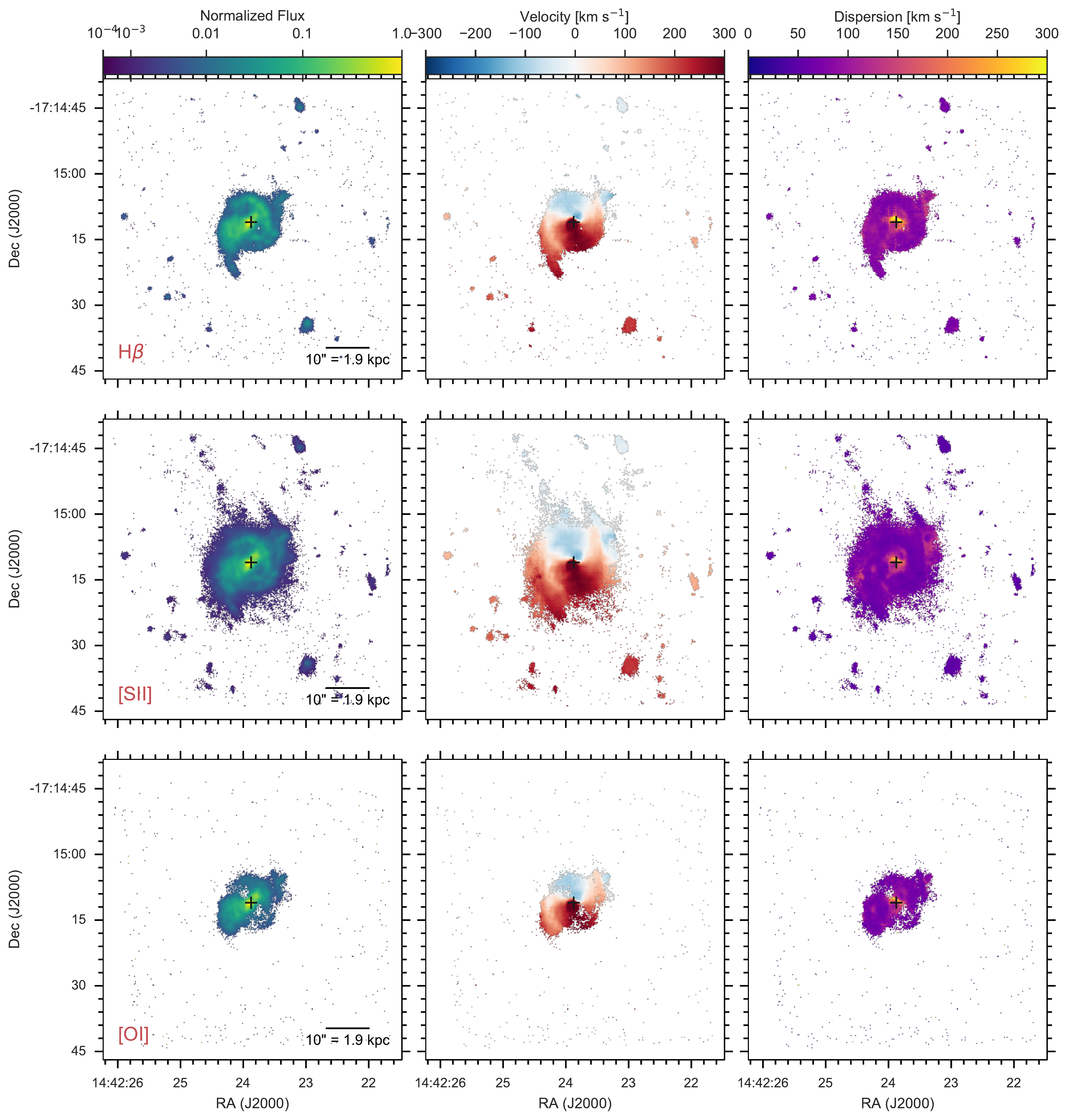}
     \caption{Same as Figure~\ref{fig:muse_1_fits} for \hbeta, [SII], and [OI]. The [SII] flux map indicates the combined flux of the $\lambda$6716 and $\lambda$6731 lines. \label{fig:muse_2_fits}}
 \end{figure*}
 
  \begin{figure*}
    \includegraphics[width=0.85\textwidth]{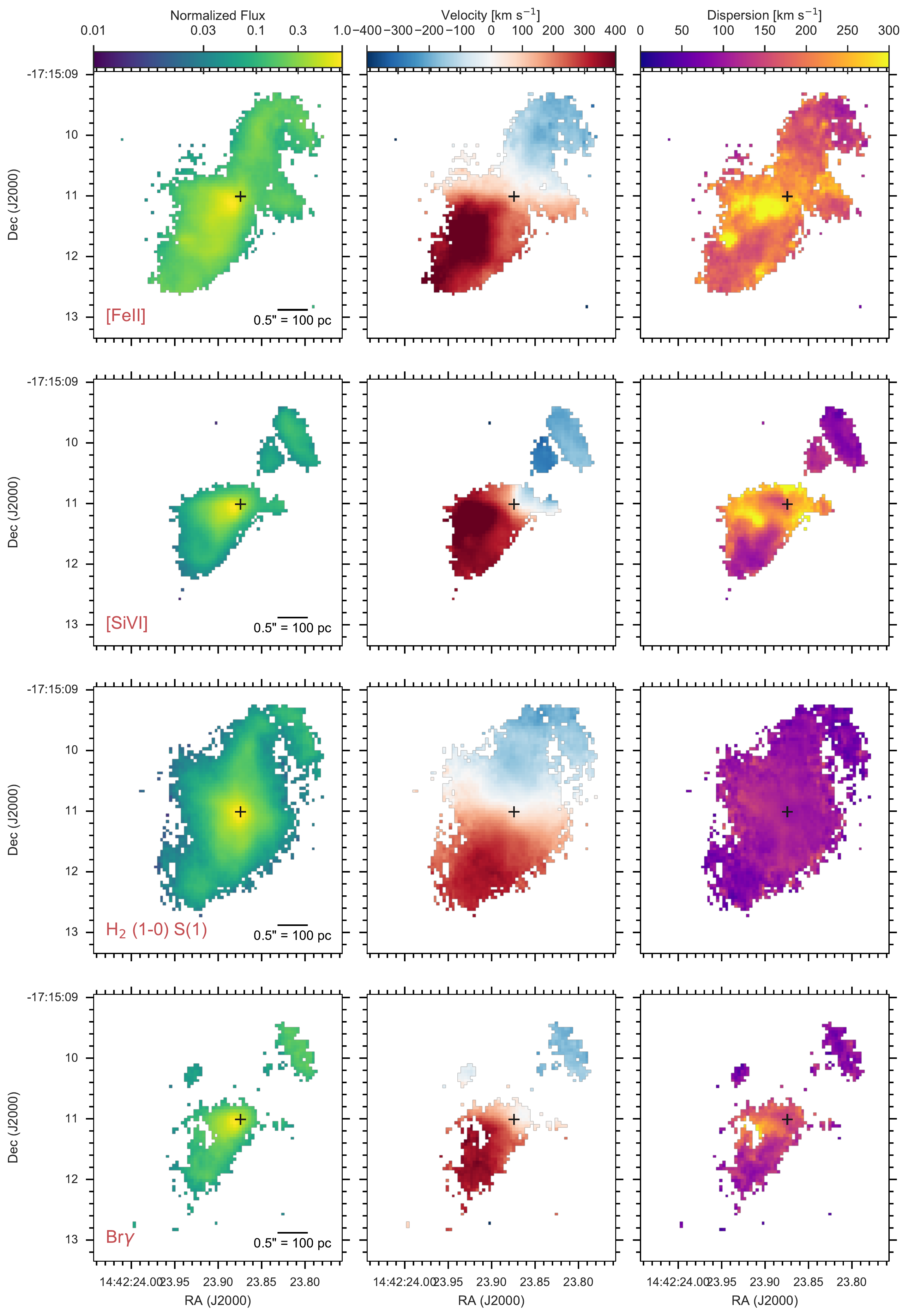}
     \caption{Single Gaussian fits to the prominent emission lines detected in the H+K band SINFONI cube. Left column shows the integrated flux normalized to the maximum flux, middle column shows the velocity, and the right column shows the velocity dispersion. Velocity dispersions have been corrected for instrumental line broadening. In all frames, North is up and East is to the left. The black cross locates the position of the AGN. \label{fig:sinfoni_fits}}
 \end{figure*}
 
 \begin{figure*}
	\includegraphics[width=\textwidth]{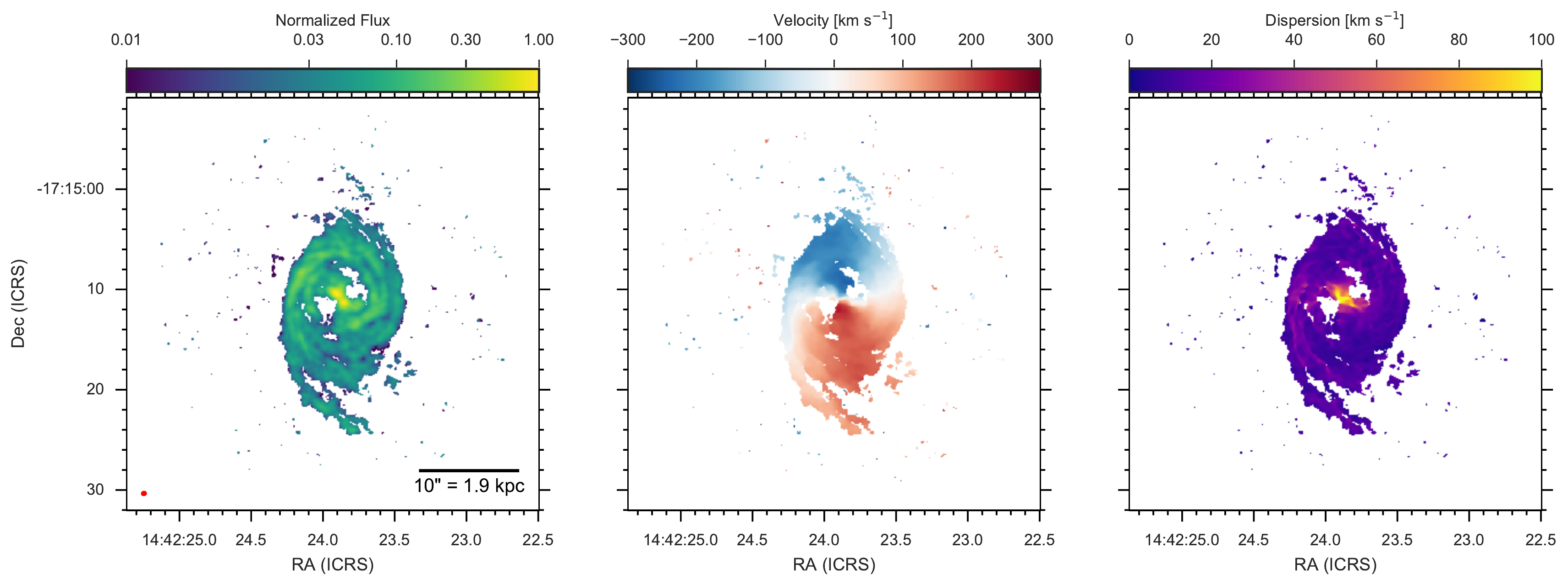}
    \caption{ \label{fig:alma_fit}Results of a single Gaussian fit to the ALMA CO 2-1 data. Left, middle, and right columns show the normalized integrated flux, velocity, and velocity dispersion. The small red ellipse in the bottom right corner indicates the size of the beam.}
    
 %The top row plots the entire ALMA FOV while the bottom row zooms in to the central 10x10\arcsec. The black contours on the integrated flux panel in the bottom row plot the location of the 1.3 mm continuum and denote the 2$\sigma$, 5$\sigma$, 10$\sigma$, 20$\sigma$, and maximum values of continuum map while the crosses show the location of the AGN based on VLBI observations. The red circles in the bottom left corner of the flux maps indicate the beam size.}
   
\end{figure*}

\section{Electron Densities}\label{sec:electron_density}
An important diagnostic IFU studies can provide is the electron density ($n_{\rm e}$) of the ionized gas. The electron density is needed in particular for the determination of mass outflow rates for both star formation and AGN driven outflows. Historically, these have been measured from single spectra using the [SII]$\lambda\lambda$6716,6731 line ratio with high uncertainties or assumed based on local studies with usual estimates of $n_{\rm e}<100\,\rm{cm}^{-3}$. Recently, IFU studies have begun to spatially resolve the electron density distribution within individual galaxies at low redshift \citep[e.g.][]{Kakkad:2018aa,Nascimento:2019aa}. At high redshift, stacks of the nuclear spectra from IFU surveys have produced estimates of the electron density for both HII regions and broad outflows \citep{Forster-Schreiber:2019aa}. These studies have all indicated that the assumed electron densities have been underestimated which in turn can greatly overestimate the mass outflow rate.

%For NGC 5728 with MUSE, we can easily determine the electron density spatial distribution from our emission line fitting of the [SII] doublet, shown in Figure~\ref{fig:muse_2_fits}. In our calculation, we assume a gas temperature of $10^4$ K and use the \textsc{pyneb}\footnote{\url{http://www.iac.es/proyecto/PyNeb/}} Python package to convert from the [SII] line ratio to electron density. Figure~\ref{fig:electron_density} shows the electron densities for every spaxel in the nuclear region of NGC 5728.

We use three independent methods to determine the electron density in the circumnuclear region of NGC 5728. The first and most popular method in the literature uses the [SII] doublet ratio as the density diagnostic due to the weak dependence on electron temperature. With a critical density around 1000 cm$^{-3}$ the [SII] line ratio provides reliable electron densities within the range of 50--2000 cm$^{-3}$. Above or below these densities however, the [SII] doublet ratio saturates and becomes unresponsive to higher or lower densities. 

The second method, introduced by \citet{Holt:2011aa} and used extensively in \citet{Rose:2018aa} and \citet{Santoro:2018aa}, incorporates the ratios of four auroral or trans-auroral emission lines: the [OII] doublets $\lambda\lambda$3726,3729 and $\lambda\lambda$7319,7330 and the [SII] doublets $\lambda\lambda$4069,4076 and $\lambda\lambda$6716,6731. This method constructs line ratios from the total flux between the doublets instead of the flux in individual lines and is therefore less susceptible to degeneracies in decomposing the doublets. The large wavelength range further provides an estimate of the gas reddening. As shown by \citet{Holt:2011aa}, this diagnostic is also sensitive to a much larger range of electron densities than the single [SII]$\lambda\lambda$6716,6731 line ratio (10$^{2}$--10$^{6.5}$ cm$^{-3}$). However, both [OII] doublets and the [SII]$\lambda\lambda$4069, 4076 doublet are usually weak and to simultaneously measure all four doublets requires either multiple spectra or a spectrum that covers a large wavelength range. 

The third method was developed in \citet[][hereafter BN19]{Baron:2019aa} and we provide a brief summary here. Instead of using the weak auroral and trans-auroral lines, they calibrated a method of measuring $n_{e}$ using the strong [OIII], \halpha, \hbeta, and [NII] lines under the assumption that the gas is photoionized by an AGN. They found that the [OIII]/\hbeta{} and [NII]/\halpha{} line ratios can provide a reliable measure of the ionization parameter following a simple equation which can then be converted into an electron density assuming the radial distance to the AGN is known. 

\begin{figure*}
	\includegraphics[width=\textwidth]{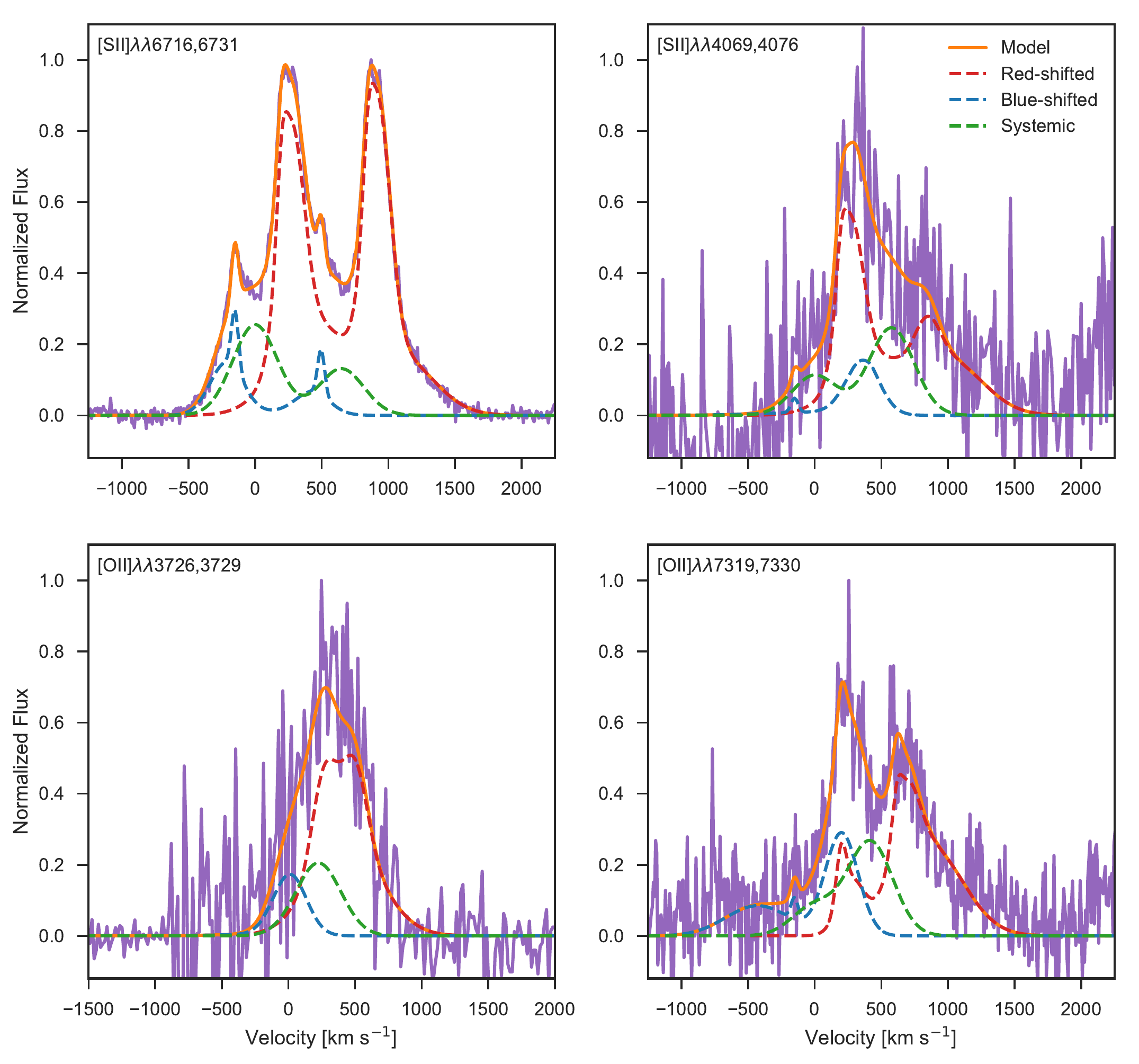}
	\caption{\label{fig:trans-auroral}X-Shooter spectrum cutouts (purple) around the four trans-auroral doublets [SII]$\lambda\lambda$6716,6731 (top left); [SII]$\lambda\lambda$4069,4076 (top right); [OII]$\lambda\lambda$3726,3729 (bottom left); and [OII]$\lambda\lambda$7319,7330 (bottom right). Overplotted are the best fit model (orange) and its red-shifted (red, dashed), blue-shifted (blue, dashed), and systemic (green, dashed) components. Spectra have been normalised to the maximum value of the cutout and the x-axis converted to velocities relative to the blue component of the doublet.}
\end{figure*}

MUSE does not extend to short enough wavelengths for the trans-auroral analysis, but as part of the LLAMA program, we obtained X-Shooter IFU spectra for all of the targets. For NGC 5728, we detected all four of the auroral or trans-auroral doublets in the spectrum extracted from a central 1.8\arcsec{} box aperture. In Figure~\ref{fig:trans-auroral}, we show cutouts of the X-Shooter spectrum around the four doublets. Unfortunately the [OII]$\lambda\lambda$3726,3729 doublet is next to the edge of one of the UVB echelle orders which caused the increased noise blue-ward of the doublet. 

To determine the necessary kinematic components to model the doublets, we first fit the [OIII]$\lambda\lambda$4959,5007 
emission lines to use as a robust template for the presence, location, and width of the existing kinematic components comprising the ionized gas as we recently did for this same galaxy in \citet{Shimizu:2018aa}. In fitting the [OIII] doublet, we fixed the velocity and dispersion of the components to be the same between each of the lines and fixed the intensity ratio (5007/4959) to 3, the theoretical value. The aperture used to produce the X-Shooter spectrum encompasses both sides of the outflow so we used a combination of red and blue-shifted Gaussian components in our model. To properly reproduce the [OIII] line profile, we found we needed 6 kinematic components, 3 red-shifted and 3 blue-shifted.

%To properly fit the [OIII] lines, we found we needed a total of 6 components, 3 red-shifted and 3 blue-shifted with the following velocities and dispersions:
%
%-- A very narrow component with velocities of $198\pm1$ and $-153\pm1$ \kms{} and dispersions of $37\pm2$ and $26\pm1$ \kms{} respectively. 
%
%-- A narrow component with velocities of $271\pm1$ and $-215\pm3$ \kms{} and dispersions of $93\pm1$ and $124\pm2$ \kms.
%
%-- A high velocity and moderately broad component with velocities of $430\pm3$ and $-430\pm44$ \kms and dispersions of $257\pm2$ and $238\pm28$ \kms. 

In fitting the auroral and trans-auroral lines, we fixed the velocity and dispersions of all of these components but left the intensities to vary. However, we found while fitting the [SII]$\lambda\lambda$6716,6731 doublet that we needed to include a seventh component with 0 \kms{} velocity and $168\pm7$ \kms{} dispersion. The remaining 3 doublets were then well-fit with these 7 components. Figure~\ref{fig:trans-auroral} shows the best fit model for each of the doublets as well as the combined red-shifted, blue-shifted, and systemic components of the model. 

While the fits overall are satisfactory, the low S/N and relatively small separation of three of the doublets creates strong degeneracies between the amplitudes of the individual components. However, as mentioned previously, the electron density measurement from the trans-auroral lines is only dependent on the total flux of each of the doublets. Therefore, for NGC 5728 the combined auroral and trans-auroral lines can provide a robust estimate of the \textit{flux-weighted} electron density but cannot provide separate electron densities for each of the kinematic components. 

\begin{figure}
	\includegraphics[width=\columnwidth]{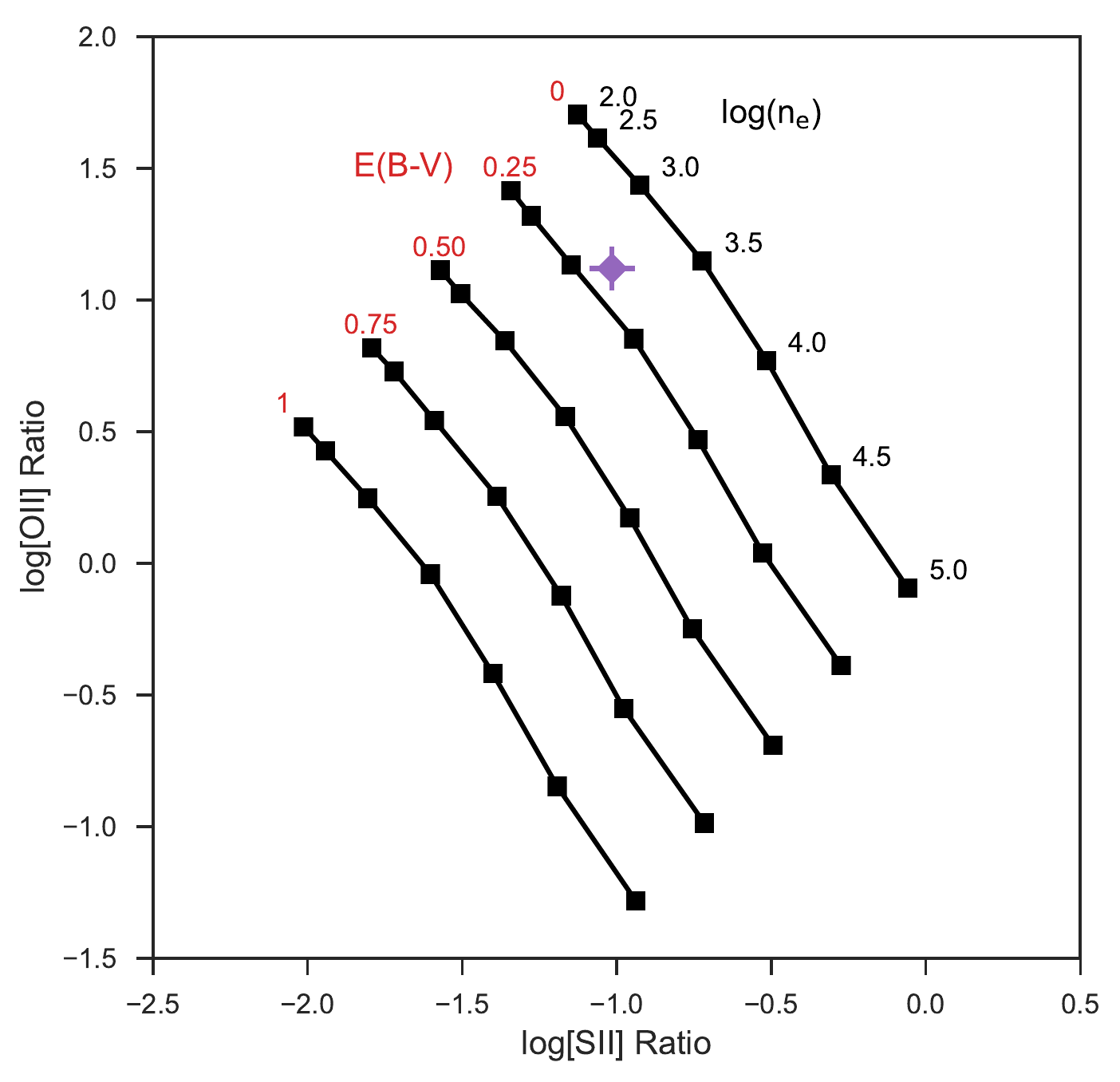}
	\caption{\label{fig:trans-auroral-diagram}Trans-auroral line ratio diagram showing the location of the [SII](4069+4076/6716+6731) and [OII](3726+3729/7319+7330) ratios for NGC 5728 (purple point). The black lines with points plot the values generated from a grid of \textsc{CLOUDY} models with a range of $\log n_{\rm e}$ and $E(B-V)$. }
\end{figure}

Figure~\ref{fig:trans-auroral-diagram} plots the location of the [SII] and [OII] doublet ratios along with the expected values for a grid of models with different $E(B-V)$ and $n_{\rm e}$. The models are the same as those presented in \citet{Rose:2018aa} and were produced using the \textsc{cloudy} photoionization code under the assumptions of solar metallicity gas photoionized by a central source with a power law continuum ($\alpha=-1.5$ for $F_{\nu} \propto \nu^{\alpha}$). They varied the electron density of the gas to model intrinsic trans-auroral line ratios and then applied a \citet{Calzetti:2000fk} reddening law for a range of $E(B-V)$. 

Using 2D spline interpolation, we estimated the best-fit $\log\,n_{\rm e}$ and $E(B-V)$ for NGC 5728 assuming they vary smoothly as a function of the trans-auroral line ratios. Taking into account only the uncertainties on the line ratios, we estimate a flux-weighted $\log n_{\rm e}=3.19\pm0.12$ (1550 cm$^{-3}$) and $E(B-V)=0.18\pm0.05$. 

Within the same X-Shooter spectrum, we then fit the \halpha+[NII] complex and \hbeta{} line profile using the same kinematic components to determine $n_{e}$ via the BN19 method. As a comparison to the trans-auroral estimate, we only calculated [OIII]/\hbeta{} and [NII]/\halpha{} line ratios for the total line profile. From the line ratios, we measure a $\log U = -2.59 \pm 0.11$. The biggest uncertainty here is what radial position to choose to convert $\log U$ into an electron density. We chose 171 pc which corresponds to half the size of the X-Shooter aperture. We find $\log n_{e} = 3.20$ which agrees well with the value from the auroral and trans-auroral lines.  

We can compare this to using only the [SII]$\lambda\lambda$6716,6731 line ratio. We find an [SII] 6716/6731 ratio of $0.65\pm0.34$. Using the \textsc{pyneb}\footnote{\url{http://www.iac.es/proyecto/PyNeb/}} software package and assuming an electron temperature of 10$^{4}$ K, this translates to a $\log n_{\rm e} = 3.36\pm0.95$ (2300 cm$^{-3}$). Clearly the large amount of blending between the [SII] doublet lines causes severe degeneracies and leads to almost a factor of ten uncertainty even though the value agrees quite well with the trans-auroral value.

\begin{figure}
	\includegraphics[width=\columnwidth]{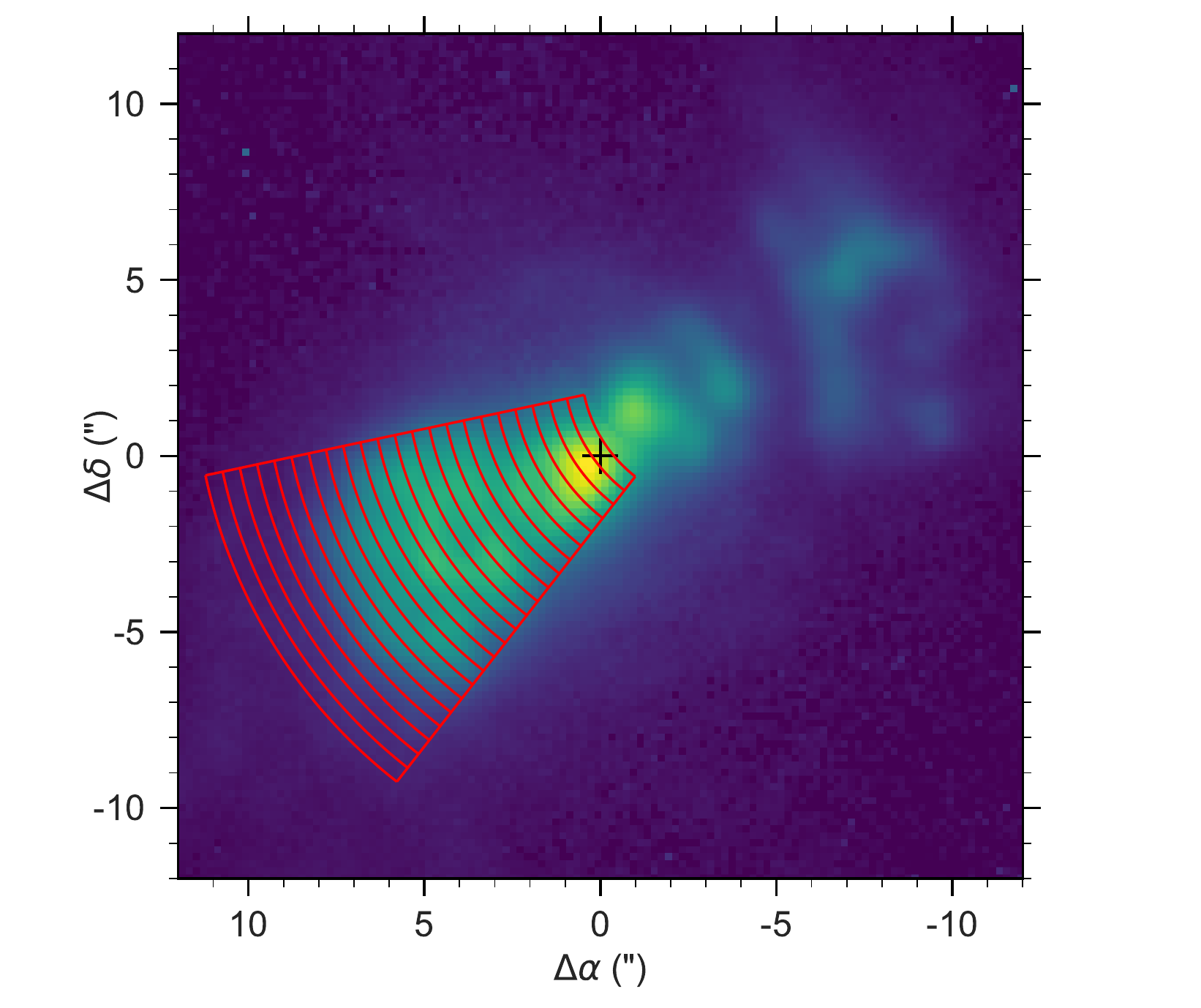}
	\caption{\label{fig:muse_annuli}[OIII]$\lambda$5007 flux map of the central 12\arcsec{} from the MUSE cube. Overlaid are the locations of the annular apertures used to measure the properties of the NLR. The black cross shows the location of the AGN.}
\end{figure}

We also compared the X-Shooter derived $n_{\rm e}$ with ones obtained from the MUSE cube based on both the [SII] 6716/6731 ratio and the BN19 method. We measured the line fluxes using annular slices along a PA of -65\degr{} West of North, ie along the bicone axis. The annular slices spanned angles +/- 20\degr{} away from the centre, had a width of 0.5\arcsec, and ranged from 0 to 11.5\arcsec{} away from the AGN. We chose to only use the redshifted SE side of the NLR since it is most unaffected by extinction and has higher S/N. Figure~\ref{fig:muse_annuli} plots the annular apertures on top of the [OIII]$\lambda$5007 flux map from our single Gaussian fits to the MUSE cube and shows how the annuli cover the majority of the biconical structure of the NLR. 

Spectra were extracted from each of the apertures and the [OIII] doublet was again fit first to determine the necessary number of kinematic components in the model since the spatial location, aperture size, spectral resolution, and S/N is different from the X-Shooter spectrum. The \hbeta, \halpha+[NII] complex, and [SII] doublet were then fit with the same components.

\begin{figure}
	\includegraphics[width=\columnwidth]{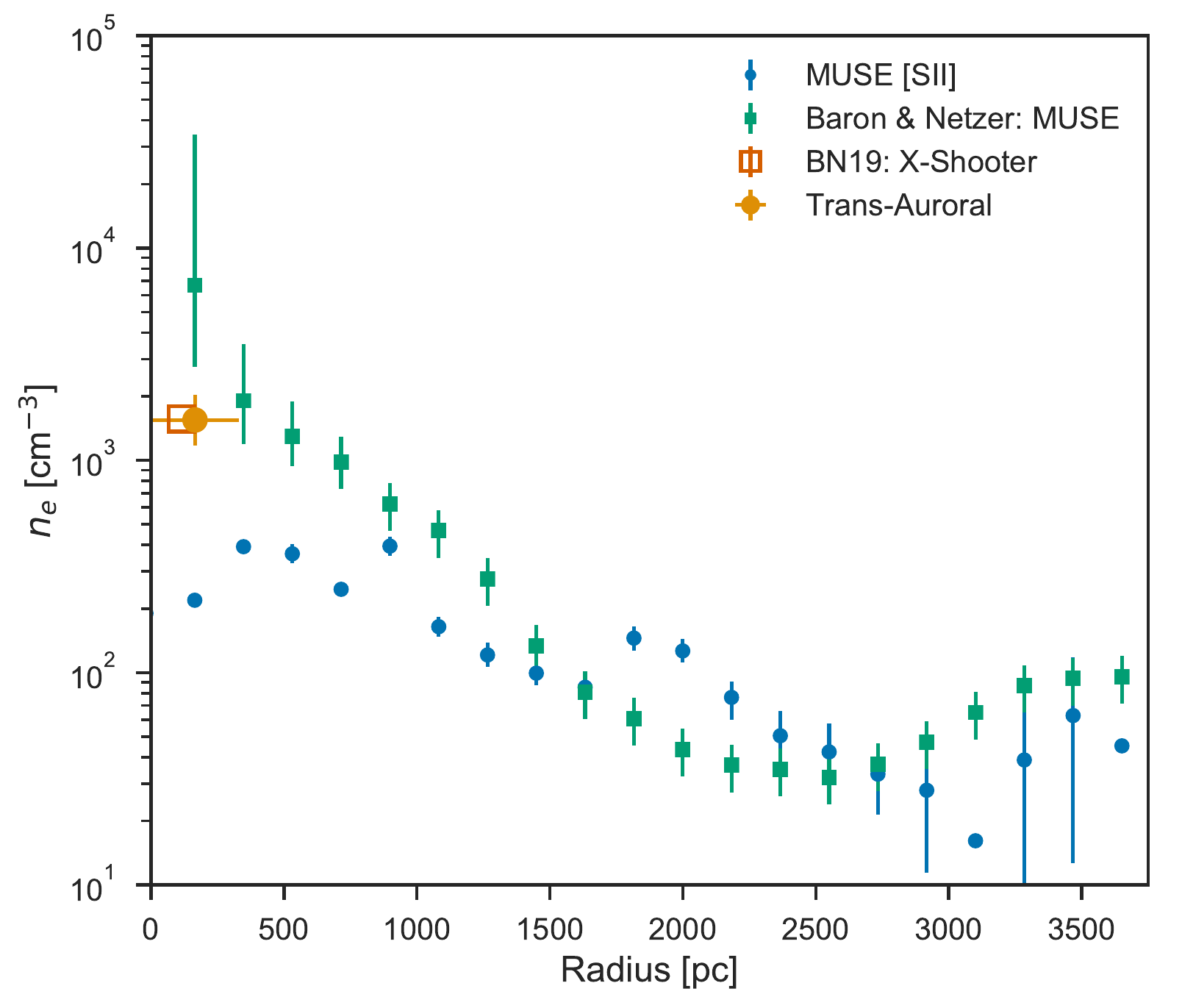}
	\caption{\label{fig:radial_ne}Radial profile of $\log\,n_{e}$ along the redshifted side of the NLR axis and within apertures shown in Figure~\ref{fig:muse_annuli}. The blue solid points plot the measurement using Gaussian fitting of the [SII] doublet emission line profile while the green squares plot the measurement based on optical line ratios and the method from \citet{Baron:2019aa}. The orange point shows the $n_{e}$ measurement from our trans-auroral emission line analysis and the open red square shows the BN19 measurement from the X-Shooter optical line ratios.}
\end{figure}

Figure~\ref{fig:radial_ne} plots $n_{\rm e}$ as a function of radius as determined by the [SII] 6716/6731 ratio and the BN19 method from the MUSE cube. Also plotted in the centre is our estimate from the X-Shooter spectrum. Inside of 1kpc, there is a strong increase of $n_{e}$ based on BN19, whereas the [SII] based $n_{e}$ slightly decreases toward the centre. This is due to the strong blending of the [SII] 6716/6731 doublet in the centre where the velocity gradient is large and there are multiple kinematic components. Therefore a robust [SII] doublet measurement of the electron density becomes very difficult without high spectral resolution. The BN19 based values are also very well aligned with the auroral and trans-auroral based value which is the most robust estimate. It's clear then that within 1 kpc, $n_{e} \approx 1000$ cm$^{-3}$, therefore for all later analysis we will use the auroral and trans-auroral value as the electron density of the gas. At larger radii, we note there is better consistency between the BN19 and [SII] doublet methods but still some discrepancies which could be explained by the different emission regions for the various lines. It's possible that while [OIII] and \halpha{} are being emitted throughout a whole ionized cloud, [SII] instead is only emitted at the ionization front. 

%There does seem to be a discrepancy between the MUSE derived densities and the trans-auroral derived density in the central region. We find a MUSE derived density for the very central aperture of $\sim$200 cm$^{-3}$ compared to 1500 cm$^{-3}$ from the trans-auroral lines. The [SII] lines in the central few spectra from MUSE however are highly blended leading to strong degeneracies between the amplitudes of the various kinematic components contained in the model. We further tested a simpler method by using the peaks of the [SII] lines to calculate the electron density and these are indicated as a dashed line in Figure~\ref{fig:radial_ne}. While the values in the centre increased to $\sim$600 cm$^{-3}$,  they still do not reach the trans-auroral value. This, combined with the large uncertainty on $n_{\rm e}$ from [SII] from X-Shooter, seems to suggest that [SII] measurements of $n_{e}$ can be unreliable in particular when a large number of kinematic components are blended together. The trans-auroral measurement is much cleaner given that it doesn't require the individual doublet lines to be decomposed, but rather just the total doublet flux. Finally, we do note that the $n_{e}$ measurements at larger radii from MUSE are more robust because the [SII] there are fewer kinematic components and the dispersions are lower allowing for better decomposition. Therefore, for the immediate central region, we will choose the trans-auroral $n_{e}$ measurement but feel safe in using the MUSE [SII] measurements at larger radii.

%%%%%%%%%%%%%%%%%%%%%%%%%%%%%%%%%%%%%%%%%%%%%%%%%%

% Don't change these lines
\bsp	% typesetting comment
\label{lastpage}
\end{document}